\documentclass[pdflatex,sn-mathphys]{sn-jnl}% Math and Physical Sciences Reference Style
\jyear{2022}%
\usepackage[section]{placeins}
\theoremstyle{thmstyleone}%
\theoremstyle{thmstyletwo}%

\theoremstyle{thmstylethree}%

\begin{document}

\title[\empty]{Spiral Sweeping Search for Smart Evaders}


\author*[1]{\fnm{Roee M.} \sur{Francos}}\email{roee.francos@cs.technion.ac.il}

\author[1]{\fnm{Alfred M.} \sur{Bruckstein}}\email{alfred.bruckstein@cs.technion.ac.il}

\affil*[1]{\orgdiv{Faculty
of Computer Science}, \orgname{Technion- Israel Institute of Technology}, \orgaddress{ \city{Haifa}, \postcode{320003}, \country{Israel}}}


\abstract{Consider a given planar circular region, in which there is an unknown number of smart mobile evaders. We wish to detect evaders using a line formation of sweeping agents whose total sensing length is predetermined. We propose procedures for designing spiral sweeping protocols that ensure the successful completion of the task, thus deriving conditions on the sweeping speed of the linear formation and its path. Successful completion of the task implies that evaders with a given limit on their speed cannot escape the sweeping agents.
A simpler task for the sweeping formation is the confinement of evaders to a desired region, such as their original domain. The feasibility of completing these tasks depends on geometric and dynamic constraints that impose a lower bound on the speed that the sweeping agents must have. This critical speed is derived to ensure the satisfaction of the confinement task. Increasing the speed above the lower bound enables the sweepers to complete the search task as well. We develop two spiral line formation search processes for smart evaders, that address current limitations in search against smart evaders. Additionally, we present a quantitative and qualitative comparison analysis between the total search time of circular line formation sweep processes and spiral line formation processes. We evaluate the different strategies by using two metrics, total search time and the minimal critical speed required for a successful search.}




\maketitle

\section{Introduction}\label{sec1}
The objective of this research is to develop efficient search policies for smart evaders. The proposed search policies are performed by a line formation of sweeping agents, or alternatively by a single agent with an equivalent sensing diameter. The formation's agents are referred to as sweepers. The goal of the sweepers is to ensure that all smart evaders that are originally inside a given circular region of radius $R_0$ are detected. The number of smart evaders and their precise locations are unknown to the sweepers. The only information that is known to the sweepers is the maximal speed of the evaders, $V_T$. The region where evaders might be located is called the evader region. The line formation of sweepers moves at a speed $V_s > V_T$ and detects evaders by using sensors with a combined length of $2r$. The search protocols may also be performed by a single agent with a line sensor of length $2r$. A line sensor with a length of $2r$ is has a rectangular shape with zero width and a length of $2r$. 

Since the original circular domain does not contain boundaries, evaders attempt to move as far as possible from the original domain that is being scanned by the line formation in order to escape detection. Hence, the goal of the evaders is avoid being detected and the goal of the line formation of sweepers is to ensure the detection of all evaders. Therefore, in this work we propose search strategies that guarantee detection of all smart evaders in the region. regardless of the trajectories that the smart evaders choose to implement. Smart evaders are agents with superior sensing and planning capabilities compared to the line formation's agents that may attempt to devise a plan that allows some evaders to escape the region, thus resulting in the failure of the line formation's objective. 

Implementing a search policy that guarantees detection of all smart evaders imposes a requirement on the minimal speed of the formation's sweepers. This speed is referred to as the critical speed, and depends on the formation's trajectory. Evaders are detected when a sweeper's sensor intersects their position. Fig. $1$ presents an illustration of the line formation's sensor. The developed protocols can be executed over a planar two dimensional region over which the formation moves or alternatively in a three dimensional domain where the formation flies over an area containing evaders. The analysis of both scenarios is equivalent.

The search protocols can be viewed as a $2$ dimensional search in which the actual agents travel on a plane or as a $3$ dimensional search where the sweepers are drone like agents which fly over the evader region. The analysis of $2D$ and $3D$ sweep protocols and of sweep protocols that are carried out by a line formation of sweepers or by a single sweeper (with equivalent sensing capabilities as the formation) are exactly the same. By assumption, the evader region has no obstacles.

\begin{figure}[ht]
\noindent \centering{}\includegraphics[width=0.4in,height =1.25in]{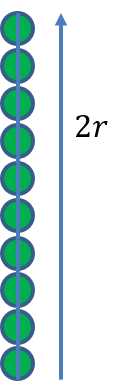} \caption{Line formation of sweeping agents that has a combined sensor diameter length of $2r$. The line formation's speed, $V_s$, is measured with respect to the formation's center.}
\label{Fig1Label}
\end{figure}

There may be two objectives that the line formation of agents may attempt to accomplish. The first is the confinement task and the second is the complete cleaning or detection task. The goal of the line formation of sweepers executing the confinement task is to prevent all evaders from leaving their original domain and escaping the line formation's agents without being detected. The result of a successful confinement task is keeping the evader region's size constant after each sweep around the region.

The ability to successfully complete the confinement task depends on a lower bound on the speed that the line formation's agents must have. This lower bound is named the critical speed and it depends on the applied search protocol. If the formation's agents move at a speed below the critical speed they will not be able to complete the confinement task.
If the line formation of sweepers speed exceeds the critical speed, the sweepers have the ability to also succeed in the complete detection task. Succeeding in the complete detection task implies that after the completion of a full sweep around the evader region, the evader region size decreases, until eventually its area is reduced to zero and all smart evaders are located.

The two metrics we use to evaluate a search protocol are the minimal sweeper’s speed required to complete the confinement task and the search time needed to detect all smart evaders in the region. The obtained results depend on the radius of the initial evader region, the evader's maximal speed and the sensing range of the line formation of sweepers.

\textbf{Contributions.} We present a full theoretical analysis that seeks to find optimal trajectories, critical speeds, and minimal sweep times for a line formation of sweepers that must ensure detection of all smart evaders originally situated inside a given circular domain. The initial circular domain that contains the evaders has no boundaries, hence the smart evaders attempt to devise a plan that will allow some of them to move out of the region that is searched by the sweepers and escape without being detected. 
\begin{itemize}
\item Theoretical analysis is provided by considering two novel line formation spiral sweep protocols:
\begin{itemize}
        \item  Drifting spiral sweep protocol 
        \item Improved spiral sweep protocol
\end{itemize}
\item For each sweep protocol, the critical speed required for the line formation of sweepers to succeed in confining the evaders is derived. Moving at speeds above the critical speed allows the line formation of sweepers to decrease the evader region. We develop explicit formulas that allow to compute the number of required sweeps, and the time, to reduce the evader region to be within a circle having a radius less than the line formation's sensing range. 
\item We propose a modification to the sweep protocols when the evader region becomes smaller than the line formation’s sensing range that allow to detect all evaders in the region, since as proved in the paper, a spiral only protocol cannot ensure the detection of all smart evaders.
\item The improved spiral sweep protocol allows to sweep the region at a nearly theoretically optimal speed.
\item The improved spiral sweep protocol ensures detection of all smart evaders at a fraction of the time compared to previous approaches, as can be seen in Fig. $14$.
\item Figures that show results from empirical simulations performed with Matlab and NetLogo are embedded in the manuscript.
\item A video attachment that graphically illustrates the dynamic evolution of the evader region and the areas detected by the line formation throughout the progression of the sweep protocols is provided as well.
\end{itemize}

\textbf{Comparison To Related Research}

This article provides an extension and a considerable improvement to our previous work \cite{francos2019search}, hence for a detailed discussion and overview regarding related research we refer the interested reader to \cite{francos2019search} and mention in this article only the most closely related work. In \cite{francos2019search}, the confinement and cleaning tasks for a line formation of agents or alternatively for a single agent with a linear sensor are analyzed. Several methods are proposed on how to determine the minimal speed for a circularly sweeping agent, in order to shrink the evader region within a circle with a smaller radius than the searcher's sensor length. The results show that this speed equals more than twice the theoretical lower bound. Furthermore, a proof that a single agent or a line formation of agents with an equivalent sensing range, employing a circular search pattern around the evader region cannot completely cleaning it without modifying the search pattern is provided. Lastly, the paper describes a modification to the trajectory of the sweepers that is performed once the evader region is bounded in a circular region with a smaller radius than half the formation's sensing range, and allows to clean the region from all evaders. 

An additional closely related work to ours is \cite{mcgee2006guaranteed} which also investigates search protocols against smart evaders. Contrary to the line sensor used by the sweeping line formation in our work, searchers in \cite{mcgee2006guaranteed} are equipped with disk shaped sensors having a radius of $r$ to detect evaders. In this work, we provide a comprehensive analysis on the time it takes to detect all evaders in the region, whereas \cite{mcgee2006guaranteed} only calculates the radius of the maximal circle that can be successfully searched and does not provide an algorithm on how to search the entire region or the time it takes to complete it. In \cite{hew2015linear}, a related problem to ours is investigated as well, however as in  \cite{mcgee2006guaranteed}, the searcher is also equipped with a disk shaped sensor that detects all evaders located at a distance of at most $r$ from the searcher. Similarly to our setting, in \cite{tang2006non}, searchers move at a constant speed and the evaders maximal speed is known to the searchers as well. In \cite{tang2006non}, the searchers have disk shaped sensors as in \cite{mcgee2006guaranteed} and  \cite{hew2015linear}.

The methodology in \cite{tang2006non} resembles our approach in the sense that ensuring that if evaders cannot escape during the first sweep around the region, they will surely not be able to escape in following sweeps around a smaller evader region. In  \cite{tang2006non}, the time it takes to completely clear the region from evaders is not explicitly calculated as in our work, however results of a simulation showing the probability to detect an evader that executes a random walk within a certain number of time steps are presented. 

Furthermore, in contrast to our approach, in \cite{mcgee2006guaranteed} and \cite{tang2006non} the searchers do not form a linear formation and are distributed equally around the region. Moving in a linear formation allows the line formation to recover more quickly from a failure of one of the sweepers by replacing the malfunctioned sweeper with the sweeper that is closest to it and consequently requiring all sweepers that are closer to the center of the evader region to redistribute themselves equally again to a smaller linear array and continue the search protocol with a reduced sensing range.  

In this work we combine ideas from our previous work \cite{francos2019search} and from \cite{mcgee2006guaranteed} in order to develop more efficient search protocols. The developed line formation spiral protocols in this work improve the search time by an order of magnitude compared to \cite{francos2019search}, while the improved spiral sweep protocol proposed in this work reduces the required critical speed of \cite{francos2019search} to nearly half its value.

\section{Spiral Sweep Protocols}\label{sec2}
This work investigates the question of devising efficient search protocols that are carried out by a single agent or alternatively by a line formation of identical sweeping agents with an equivalent sensing range that must guarantee detection of all smart evaders originally location inside a circular region of radius $R_0$. The number of evaders and their locations is unknown to the sweeping agents, hence it is assumed that evaders can be originally located anywhere inside the evader region and that once the search protocol commences they may move out of the region at a maximal speed of $V_T$ in all direction. The sweepers move in a way that the line formation advances, most often perpendicularly to the agents' linear array with a speed of $V_s$ (measured at the center of the linear sensor). We assume that evaders move at a maximal speed of $V_T$, and do not have any maneuverability restrictions. The sweepers have two potential tasks they try to achieve. The sweepers' first goal is to ensure that no evader escapes the region being searched undetected, while keeping the size of the evader region constant after each sweep. If the line formation's speed exceeds the critical speed that allows to satisfy the confinement goal, then the goal of the formation is to implement a sweeping protocol that enables the fastest detection of all evaders in the region, by reducing to zero the evader region's area. This is achieved by iteratively reducing the radius of the circle bounding the evader region after the completion of each sweep.

Planning against smart evaders can be seen as solving the worst-case scenario of a given situation and hence the applied search protocols are preprogrammed and deterministic to ensure that the chosen search protocol allows to detect all evaders regardless of the escape plans they choose. The time it takes to completely detect all evaders naturally depends on the applied search protocol. In this work the line formation of sweepers performs spiral search protocols that track the advancing wavefront of potential evader locations. The line formation of sweepers starts the protocol when its entire sensor of length $2r$ is inside the evader region, thus allowing it to detect as many evaders as possible.

In order to compare the developed search protocols in this work and to compare them to previous methods as well, a lower bound on the line formation's speed independent of the search protocol is presented. Afterwards, we present two different spiral line formation search protocols that can be applied. For each protocol, we determine the minimal speed required by the line formation to decrease the evader region to be within a circle having a smaller radius than $2r$.

The first spiral search protocol, the drifting spiral search protocol considers only spiral sweeps. After each sweep around the evader region, a new center for the evader region is computed and the next sweep is performed with respect to that center. This new evader region center is computed as the midpoint between the highest and lowest points the evader region spread to, after a completion of a sweep around the region. A completion of a full sweep around the evader region is called a cycle or an iteration. Changing the center of the evader region after each sweep enables the sweep protocol to consist of only spiral sweeps without linear inward advancements in which no cleaning occurs, as are performed in \cite{francos2019search}. This protocol allows for a simpler search procedure that does not require any assumptions on the turning rates of the sweepers. This sweep protocol implies that after each sweep the center of the new evader region moves upwards by a positive distance. The drifting spiral search protocol yields that the minimal agent speed that ensures satisfaction of the confinement task is lower than the critical speed developed in \cite{francos2019search}, however it is still not close to the lower bound on a searcher speed, and hence is not optimal. An analytical formula that calculates the number of required scans that are needed in order to reduce the evader region to be contained in a circle with a smaller radius than $r$, is then derived. The drifting spiral search protocol enables derivation of analytical formulas for the time it takes the sweepers to reduce the evader region to be bounded by a circle of radius less than or equal to $2r$ as a function of the search parameters. 

In order to improve the drifting spiral search protocol and reduce the critical speed that enables confinement as well as decreasing the sweep time it takes to reduce the evader region to be bounded by a circle of radius less than or equal to $2r$, we present a different and improved spiral sweep protocol. This protocol does not allow for a derivation of analytical formulas for the entire search protocol, however it yields nearly optimal results. In the improved spiral protocol the center of the evader region stays fixed throughout the search. We then show that for a line formation of sweepers equipped with line sensors, a spiral sweep pattern around the evader region cannot complete the cleaning of the entire area using only spiral sweeping. This applies for both developed spiral protocols. In order to solve the problem we provide a modification for the search protocol that is applied when the evader region is bounded by a circle with a radius of less than $2r$. We then show that if the ratio between the searcher speed $V_s$ and the evader's maximal speed $V_T$ is above a certain threshold, the sweeper formation can completely clean the region performing the modified algorithm.  Illustrative simulations that demonstrate the evolution of the search protocols were generated using NetLogo software \cite{tisue2004netlogo}.

\textbf{Paper Organization.} 
Section \ref{sec3} presents a lower bound on the critical speed a line formation of sweepers is required to have to enable it to succeed in the confinement task. Section \ref{sec4} provides an analysis of the drifting spiral sweep protocol. A critical speed required for the sweeping line formation to perform a search protocol that guarantees detection of all evaders is presented along with a sweep time analysis of the proposed protocol. Section \ref{sec5} provides an analysis of the improved spiral sweep protocol. Section \ref{sec6} provides comparative analysis between the proposed spiral search protocols as well as a comparison to the circular sweep method developed in \cite{francos2019search} showing the clear advantages of the proposed protocols developed in this work.  

\section{A Universal Bound on Cleaning Rate}
\label{sec3}
In order to compare the performance of different search protocols we use two metrics:  the first one is the minimal critical speed that allows to successfully accomplish the confinement task and the second one is the time it takes to detect all evaders in the region. For the first metric, the critical speed, there exists a lower bound that is independent of the sweep protocol employed by the line formation. This implies that if the line formation of sweepers moves at a speed below this theoretical lower bound it will not be able to succeed in the confinement task regardless of the search pattern it implements.

The full proof is provided in \cite{francos2019search}, however, the main considerations and notations are provided in this section as well since in later sections we compare the critical speeds of developed search protocols to this lower bound to assess their quality.  

Denote the sweeper’s speed as $V_s$, the sensor length as $2r$, the evader region's initial radius as $R_0$ and the maximal speed of evaders as $V_T$. The lower bound on the speed is established by requiring that the maximal cleaning rate (rate of detecting evaders) is larger than the minimal expansion rate of the evader region. The resulting speed that arises from this consideration is defined as the critical speed and denoted by ${{\rm{V}}_{LB}}$.
 
\newtheorem{thm}{Theorem}
\begin{thm}
No sweeping protocol is able to successfully complete the confinement task if its speed, $V_s$, is less than,
\begin{equation}
{{\rm{V}}_{LB}} = \frac{{\pi {R_0}{V_T}}}{r}
\label{e1}    
\end{equation}
\end{thm}
For proof see \cite{francos2019search}. As the line formation progresses with its search protocol, it decreases the evader region's area. Therefore, after each sweep, the line formation sweeps around a region bounded by a circle with a decreased radius and perimeter compared to the initial perimeter. Hence, if the line formation moves with a speed that enables it to confine all evaders in the initial sweep, it will surely have sufficient speed to prevent escape of evaders during the next sweeps around a smaller evader region. 

\section{The Drifting Spiral Sweep Protocol}
\label{sec4}
At the start of every circular sweep in the circular sweep protocol developed in \cite{francos2019search}, half of the line formation's sensor is outside of the evader region. Naturally, in order for the sweeping line formation to employ a more efficient motion throughout the search protocol, the footprint of the formation's sensor that overlaps the evader region should be maximal. This is obtained by performing a spiral trajectory, in which the formation's sensor tracks the expanding evader region's wavefront, while at the same time it attempts to preserve the evader region's shape to be nearly circular. An illustration of the initial placement of a line formation of agents that employs the drifting spiral sweep protocol is presented in Fig. $2$.
\begin{figure}[ht]
\noindent \centering{}\includegraphics[width=2.3in,height =2in]{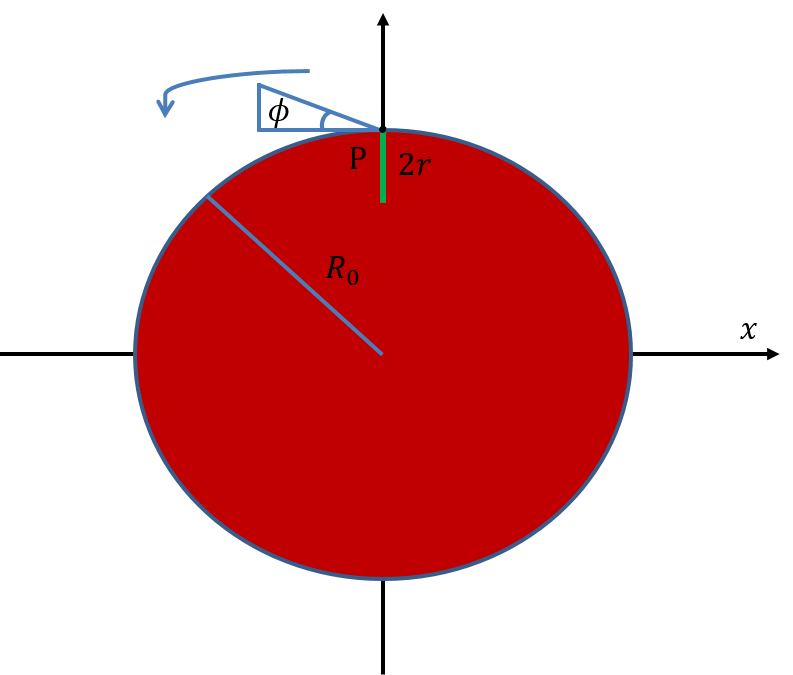} \caption{Initial placement of a line formation of sweepers implementing a spiral sweep protocol. The formation's sensor is shown in green. Red areas indicate locations where potential evaders may be located. The angle $\phi$ is the angle between the tip of the formation's sensor and the normal of the evader region. $\phi$ is an angle given by the ratio of sweeper and evader velocities.}
\label{Fig2Label}
\end{figure}
The line formation of agents has a sensor of length $2r$. We wish the sweepers to have the lowest possible critical speed, hence we propose a different search protocol. A lower critical speed enables the sweepers to scan larger evader regions successfully and enables shorter cleaning times of the evader region, given an agent's speed. The line formation of sweeping agents starts with a sensor length of $2r$ inside the evader region. If the sweeper formation's speed is above the scenario's critical speed, the sweeper formation reduces the evader region's area after completing a traversal of $2\pi$ degrees around the evader region.

The line formation of sweepers begins its spiral traversal with the upper tip of its sensor tangent to the edge of the evader region at point $P =(0,R_0)$. In order to keep its sensor tangent to the evader region, the formation must travel at angle $\phi$ to the normal of the evader region. $\phi$ is calculated by,
\begin{equation}
\sin \phi  = \frac{{{V_T}}}{{{V_s}}}
\label{e2}
\end{equation}
Thus we have,
\begin{equation}
\phi  = \arcsin \left( {\frac{{{V_T}}}{{{V_s}}}} \right)
\label{e3}
\end{equation}
This method of traveling at angle $\phi$ preserves the evader region's circular shape. The sweeper formation travels along the perimeter of the evader region. The isoperimeteric inequality states that for a given area the shape of the curve that bounds this area that has the smallest perimeter is the circle. Therefore, this method ensures that the time it takes to complete a sweep around the evader region is minimal. The equations of motion are calculated with respect to the center of the sweeper formation. 

The time it takes the formation to complete a spiral traversal around the angular region of the evader region it is responsible to scan corresponds to changing its angle $\theta$ by $2\pi$. We denote by $C\left( {{t_\theta }} \right)$ the trajectory of the center of the line formation, or alternatively the center of a single sweeper's sensor, as a function of $\theta$, the angle of the sweeper sensor with respect to the center of the evader region. The trajectory is given by,   
\begin{equation}
C\left( {{t_\theta }} \right) = \left( {{R_0} - r + {V_T}{t_\theta }} \right)\left[ {\sin \left( {\theta \left( {{t_\theta }} \right)} \right),\cos \left( {\theta \left( {{t_\theta }} \right)} \right)} \right]
\label{e4}
\end{equation}
Similarly, the trajectories of the tip of the sensor that is the furthest from the center of the evader region is given by,
\begin{equation}
U\left( {{t_\theta }} \right) = \left( {{R_0} + {V_T}{t_\theta }} \right)\left[ {\sin \left( {\theta \left( {{t_\theta }} \right)} \right),\cos \left( {\theta \left( {{t_\theta }} \right)} \right)} \right]
\label{e5}
\end{equation}
And the trajectory of the tip of the sensor that is closest to the center of the evader region is given by,
\begin{equation}
L\left( {{t_\theta }} \right) = \left( {{R_0} - 2r + {V_T}{t_\theta }} \right)\left[ {\sin \left( {\theta \left( {{t_\theta }} \right)} \right),\cos \left( {\theta \left( {{t_\theta }} \right)} \right)} \right]
\label{e6}
\end{equation}
The formation's angular speed, or the rate of change of its angle with respect to the center of the evader region, can be described as a function of $\phi$ as,
\begin{equation}
\frac{{\theta \left( {{t_\theta }} \right)}}{{d{t_\theta }}} = \frac{{{V_s}\cos \phi }}{{{R_s}({t_\theta })}} = \frac{{\sqrt {{V_s}^2 - {V_T}^2} }}{{{R_s}({t_\theta })}}
\label{e7}
\end{equation}
Where ${t_\theta }$ denotes the sweep time after the sweeper completed a traversal of angle $\theta$ around the center of the evader region. ${R_s}({t_\theta })$ is equal to the distance of $C\left( {{t_\theta }} \right)$ from the center of the evader region. The instantaneous growth rate of the distance between the formation's center and the center of the evader region is given by,
\begin{equation}
\frac{{d{R_s}( {t_\theta })}}{{d{t_\theta }}} = {V_s}\sin \phi  = {V_T}
\label{e8}
\end{equation}
Integrating the left and right sides of equation (\ref{e7}) between the initial search time and time ${t_\theta }$ yields,
\begin{equation}
\int_0^{{t_\theta }} {\dot \theta \left( \zeta  \right)} d\zeta  = \int_0^{{t_\theta }} {\frac{{\sqrt {{V_s}^2 - {V_T}^2} }}{{{V_T}\zeta  + {R_0} - r}}d} \zeta 
\label{e9}
\end{equation}
Therefore we have that,
\begin{equation}
\theta \left( {{t_\theta }} \right) = \frac{{\sqrt {{V_s}^2 - {V_T}^2} }}{{{V_T}}}\ln \left( {\frac{{{V_T}{t_\theta } + {R_0} - r}}{{{R_0} + r}}} \right)
\label{e10}
\end{equation}
Applying the exponent function to both sides of (\ref{e10}) and rearranging terms yields,
\begin{equation}
{e^{\frac{{{V_T}{t_\theta }}}{{\sqrt {{V_s}^2 - {V_T}^2} }}}} = \frac{{{V_T}{t_\theta } + {R_0} - r}}{{{R_0} - r}}
\label{e11}
\end{equation}
Thus we obtain,
\begin{equation}
{V_T}{t_\theta } = \left( {{R_0} - r} \right)\left( {{e^{\frac{{{V_T}{t_\theta }}}{{\sqrt {{V_s}^2 - {V_T}^2} }}}} - 1} \right)
\label{e12}
\end{equation}
This yields that the time it takes the sweeper formation to travel an angle of $\theta$ around the center of the evader region is,
\begin{equation}
{t_\theta } = \frac{{\left( {{R_0} - r} \right)\left( {{e^{\frac{{\theta {V_T}}}{{\sqrt {{V_s}^2 - {V_T}^2} }}}} - 1} \right)}}{{{V_T}}}
\label{e13}
\end{equation}
Consequently, the time it takes it to complete a full sweep around the region is,
\begin{equation}
{t_{2\pi }} = \frac{{\left( {{R_0} - r} \right)\left( {{e^{\frac{{2\pi {V_T}}}{{\sqrt {{V_s}^2 - {V_T}^2} }}}} - 1} \right)}}{{{V_T}}}
\label{e14}
\end{equation}
At time ${t_\theta }$ the sweeper formation detects the evaders up to the point $L\left( {{t_\theta }} \right)$ given in (\ref{e6}). From ${t_\theta }$ to ${t_{2\pi }}$ the evader region expands from  $L\left( {{t_\theta }} \right)$ in all directions with a maximal speed of $V_T$ for a time of ${t_{2\pi }}-{t_\theta }$, resulting in a spread of radius ${V_T}\left( {{t_{2\pi }} - {t_\theta }} \right)$ in all directions. Hence, the wavefront from $L\left( {{t_\theta }} \right)$ is defined by the curve, 
\begin{equation}
E\left( {\theta ,\psi } \right) = L\left( {{t_\theta }} \right) + {V_T}\left( {{t_{2\pi }} - {t_\theta }} \right)\left[ {\sin \left( \psi  \right),\cos \left( \psi  \right)} \right]
\label{e15} 
\end{equation}
for all $\psi  \in \left[ {0,2\pi } \right]$. Equation (\ref{e15}) shows the expansion from $L\left( {{t_\theta }} \right)$ at time ${t_{2\pi }}$. All the points of the wavefront are defined at ${t_{2\pi }}$
by,
\begin{equation}
\begin{array}{l}
E\left( {\theta,\psi} \right) = \\ \left( {{R_0} - 2r + {V_T}{t_\theta }} \right)\left[ {\sin \left( {\theta \left( {{t_\theta }} \right)} \right),\cos \left( {\theta \left( {{t_\theta }} \right)} \right)} \right] +  {V_T}\left( {{t_{2\pi }} - {t_\theta }} \right)\left[ {\sin \left( \psi  \right),\cos \left( \psi  \right)} \right]    
\end{array}
\label{e16}    
\end{equation}
After a traversal of $2\pi$ around the evader region, the furthest tip of the formation's sensor from the center of the evader region is located at point, 
\begin{equation}
U\left( {{t_{2\pi }}} \right) = \left( {0,{R_0} + {V_T}{t_{2\pi }}} \right)
\label{e17}
\end{equation}
The point that started to spread at time ${t_\pi}$ in the direction of the negative $y$ axis is the one that is furthest away from $U\left( {{t_{2\pi }}} \right)$. This point is described by $E\left( {\theta=\pi,\psi = \pi} \right)$,
\begin{equation}
E\left( {\theta  = \pi ,\psi  = \pi } \right) = \left( {0, - {R_0} + 2r - {V_T}{t_{2\pi }}} \right)
\label{e18}
\end{equation}
The new center point of the evader region is calculated as the midpoint between $E\left( {\theta  = \pi ,\psi  = \pi } \right)$ and $U\left( {{t_{2\pi }}} \right)$ and results in, 
\begin{equation}
\frac{{U\left( {{t_{2\pi }}} \right) + E\left( {\theta  = \pi ,\psi  = \pi } \right)}}{2} = \left( {0,r} \right)
\label{e19}
\end{equation}
This implies that after the completion of the sweep, the center of the evader region moves up by a distance of $r$. The new radius of the circle that bounds the evader region is given as half the difference between the $y$ values of $E\left({\theta  = \pi ,\psi  = \pi } \right)$ and $U\left( {{t_{2\pi }}} \right)$ and is given by,
\begin{equation}
{R_{new}} = \frac{{{y_u}\left( {{t_{2\pi }}} \right) - {y_E}\left( {\theta  = \pi ,\psi  = \pi } \right)}}{2} = {R_0} - r + {V_T}{t_{2\pi }}= \left( {{R_0} - r} \right){e^{\frac{{2\pi {V_T}}}{{\sqrt {{V_s}^2 - {V_T}^2} }}}}
\label{e20}
\end{equation}
We want to show that at time ${t_{2\pi }}$ all points from $E\left( {\theta  = 2\pi ,\psi } \right)$
are at a distance from $(0,r)$ that is less than ${R_0} - r + {V_T}{t_{2\pi }}$.
The furthest evaders are the ones that start from $L\left( {{t_\theta }} \right) $, and escape radially from the new center of the evader region located at $(0,r)$. Therefore, in order to guarantee confinement of all evaders at time ${t_{2\pi }}$ the following inequality has to be satisfied,
\begin{equation}
\begin{array}{l}
\sqrt {{{\left( {\left( {{R_0} - 2r + {V_T}{t_\theta }} \right)\sin \left( {\theta \left( {{t_\theta }} \right)} \right)} \right)}^2} + {{\left( {\left( {{R_0} - 2r + {V_T}{t_\theta }} \right)\cos \left( {\theta \left( {{t_\theta }} \right)} \right) + r} \right)}^2}} \\  + {V_T}\left( {{t_{2\pi }} - {t_\theta }} \right) \le {R_0} - r + {V_T}{t_{2\pi }}
\end{array}
\label{e21}
\end{equation}
Using simple trigonometric identities, squaring both sides of the equation and rearranging terms yields,  
\begin{equation}
{\left( {{R_0} - 2r + {V_T}{t_\theta }} \right)^2} + 2r\left( {{R_0} - 2r + {V_T}{t_\theta }} \right)\cos \left( {\theta \left( {{t_\theta }} \right)} \right) + {r^2} \le {\left( {{R_0} - r + {V_T}{t_\theta }} \right)^2}
\label{e22}
\end{equation}
Rearranging terms yields that,
\begin{equation}
2r\left( {{R_0} - 2r + {V_T}{t_\theta }} \right)\cos \left( {\theta \left( {{t_\theta }} \right)} \right) + {r^2} \le 2r\left( {{R_0} - r + {V_T}{t_\theta }} \right) - {r^2}
\label{e23}
\end{equation}
We therefore obtain that,
\begin{equation}
\cos \left( {\theta \left( {{t_\theta }} \right)} \right) \le 1
\label{e24}
\end{equation}
Since this inequality is always satisfied, it means that all evaders are bounded in a circular region of radius ${R_0} - r + {V_T}{t_{2\pi }}$. In order to successfully complete the confinement task we demand that after the sweeper formation completes its traversal around the evader region the region is bounded by a circle with a smaller or equal radius than $R_0$, hence,
\begin{equation}
{R_{new}} =\left( {{R_0} - r} \right){e^{\frac{{2\pi {V_T}}}{{\sqrt {{V_s}^2 - {V_T}^2} }}}}\le {R_0}
\label{e25}
\end{equation}
Since the logarithm function is a monotonically increasing function, applying it to both sides of (\ref{e25}) does not change the direction of the inequality. We therefore obtain,
\begin{equation}
\frac{{2\pi {V_T}}}{{\ln \left( {\frac{{{R_0}}}{{{R_0} - r}}} \right)}} \le \sqrt {{V_s}^2 - {V_T}^2} 
\label{e26}
\end{equation}
Therefore, in order to have a no escape sweep protocol the sweepers' speeds have to satisfy,
\begin{equation}
{V_S} \ge {V_T}\sqrt {\frac{{4{\pi ^2}}}{{{{\left( {\ln \left( {\frac{{{R_0}}}{{{R_0} - r}}} \right)} \right)}^2}}} + 1}
\label{e27}    
\end{equation}
When choosing the following search parameters, ${R_0}=100$,$r=10$,${V_T}=1$, we obtain that the optimal critical speed is ${V_{LB}}=31.4159$ while the spiral critical speed has a value of ${V_c}_{_{spiral}}=59.6435$. This value is lower than the circular critical speed in \cite{francos2019search} which resulted in a value of $62.8319$.
An illustrative simulation that demonstrates the confinement task when the sweeper formation employs the drifting spiral sweep protocol is presented in Fig. $3$. The sweeper line formation confines evaders to an area of the same size as initial evader region. The new evader region is an upwards shifted version of the initial evader region. 
\begin{figure}
\noindent \centering{}\includegraphics[width=1.5in,height =1.5in]{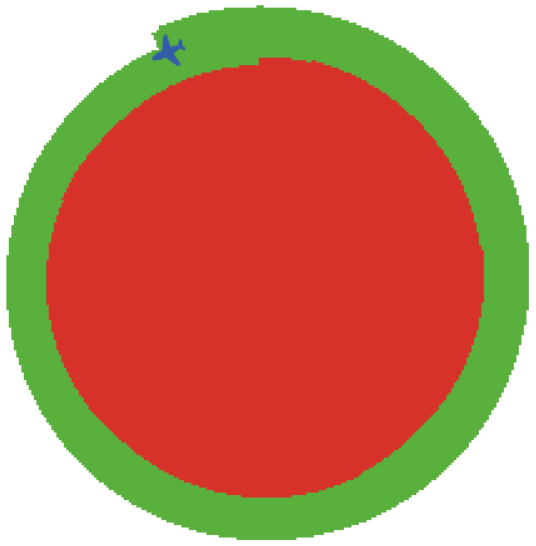} \caption{Swept areas and evader region status for a scenario where the line formation of agents successfully confines evaders to an area of the same size as initial evader region with the drifting spiral sweep protocol. The new evader region is an upwards shifted version of the initial evader region. Green areas are locations that are free from evaders and red areas indicate locations where potential evaders may still be located.}
\label{Fig3Label}
\end{figure}
In order for the sweeper formation to shrink the evader region, it must travel in a speed that is greater than the critical speed. We denote by $\Delta V$ the increment in the sweeper formation's speed that is above the critical speed. The formation's speed ${V_s}$, is therefore given by the sum of the critical speed and $\Delta V$, namely
${V_s} = {V_c} + \Delta V$.
In our setting the angle that the sweeper formation traverses in every cycle is equal to $2\pi$. Let us denote by $T_i$, the time it takes the sweeper formation to travel around a circular region of radius ${R_i}$. From (\ref{e14}) we have that $T_i$ is given by,
\begin{equation}
{T_i} = \frac{{\left( {{R_i} - r} \right)\left( {{e^{\frac{{2\pi {V_T}}}{{\sqrt {{V_s}^2 - {V_T}^2} }}}} - 1} \right)}}{{{V_T}}}
\label{e28}
\end{equation}
In the proposed line formation spiral sweep, the center of the evader region does not remain fixed. After the completion of each sweep it moves up by a positive distance. Therefore, as the sweepers progress in their search mission, the evader region shrinks while the region itself moves upwards. If the agents in the formation move at a speed greater than the critical speed for the corresponding scenario they can progress with the search task. An illustrative simulation that demonstrates the evolution of the search protocol when the sweeper formation employs the drifting spiral sweep protocol is presented in Fig. $4$. Green areas are locations that are free from evaders and red areas indicate locations where potential evaders may still be located. 
\begin{figure}
\noindent \centering{}\includegraphics[width=2.6in,height =6in]{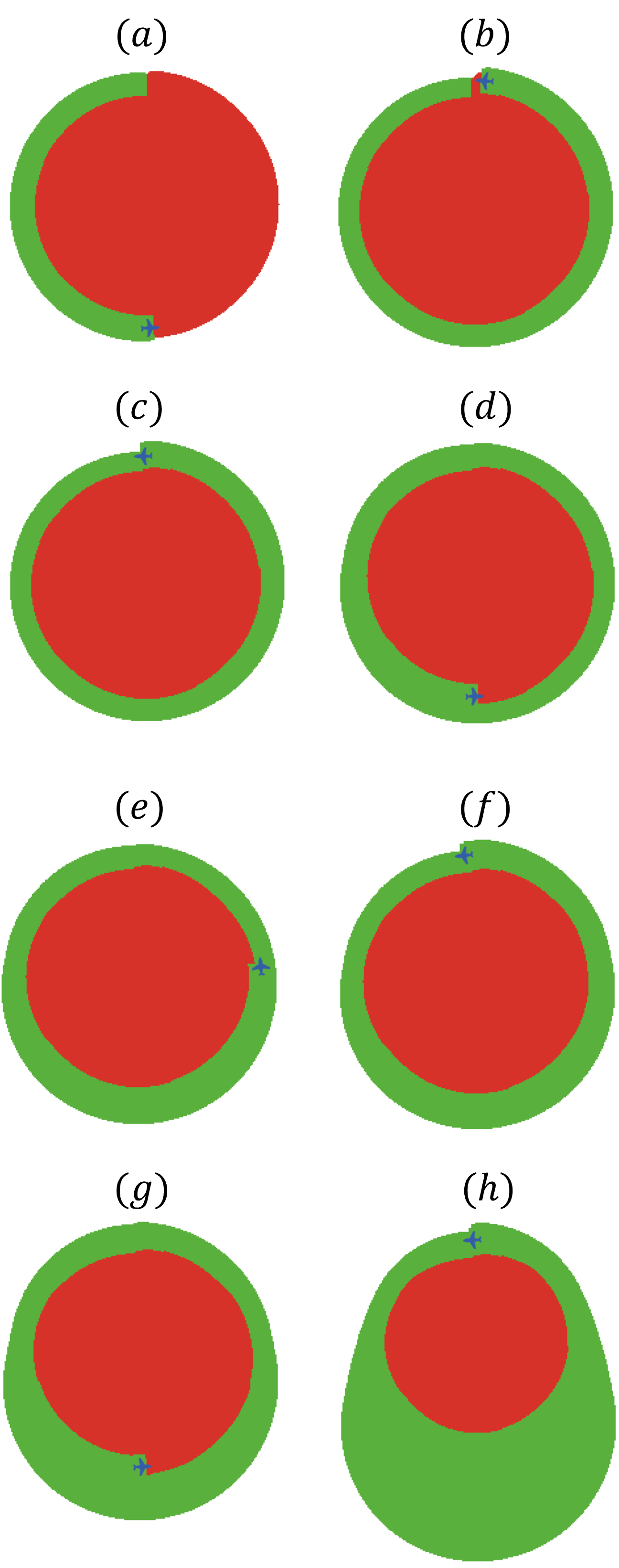} \caption{Swept areas and evader region status for different times in a scenario where the line formation of agents employs the drifting spiral sweep protocol. Green areas are locations that are free from evaders and red areas indicate locations where potential evaders may still be located. (a) - After a sweep by an angle of $\pi$. (b) - Towards the end of the first sweep. (c) - Beginning of the second sweep. (d) - Halfway through the second sweep. (e) - After a sweep of $\frac{3\pi}{2}$ in the second sweep. (f) - Beginning of the second sweep. (g) - Beginning of the third sweep. (h) - Beginning of the tenth sweep.}
\label{Fig4Label}
\end{figure}
After each sweep around the region, the evader region is bounded by a circle whose radius can be analytically calculated. From (\ref{e20}) We have that,
\begin{equation}
{R_{i + 1}} = {R_i}{e^{\frac{{2\pi {V_T}}}{{\sqrt {{V_s}^2 - {V_T}^2} }}}} - r{e^{\frac{{2\pi {V_T}}}{{\sqrt {{V_s}^2 - {V_T}^2} }}}}
\label{e29}
\end{equation}
We denote by $\widetilde{R}_i=R_i - r$. Using $\widetilde {R}_i$ instead of $R_i$, yields equations that have the same structure as the equations that were developed for the circular sweep protocol in \cite{francos2019search}. This allows to solve the resulting spiral sweep protocol's equations with the same methodology along with the appropriate change of coefficients. Thus (\ref{e29}) reduces to,
\begin{equation}
\widetilde{R}_{i + 1} = {c_2}\widetilde{R}_i + {c_1}
\label{e30}
\end{equation}
Where we denote the equation's coefficients by,
\begin{equation}
{c_1} =  - r , \hspace{1mm} {c_2} = {e^{\frac{{2\pi {V_T}}}{{\sqrt {{V_s}^2 - {V_T}^2} }}}}
\label{e31}
\end{equation}
The formula for the number of iterations it takes the sweeper formation to reduce the evader region to be bounded by a circle with a radius of $\widehat{R}_N=2r$, is developed in Appendix $A$ of \cite{francos2019search} and is given by,
\begin{equation}
N = \left\lceil {\frac{{\ln \left({\frac{{\widehat{R}_N - \frac{{{c_1}}}{{1 - {c_2}}}}}{{{R_0} - \frac{{{c_1}}}{{1 - {c_2}}}}}} \right)}}{{\ln {c_2}}}} \right\rceil
\label{e32}
\end{equation}
The exact value of $R_N$ can only be calculated after the number of sweeps around the region, $N$, is calculated. Therefore, we use $\widehat{R}_N=2r$ as an estimate of the true $R_N$ in order to calculate $N$. $R_N$, the actual radius of the circle that bounds the evader region when it becomes smaller or equal to $2r$, is calculated in Appendix $B$ of \cite{francos2019search}. The precise calculation of $R_N$ is important for the end game of the cleaning protocol as is discussed later in this section. The last sweep takes place when the evader region is bounded by a circle of radius, $\widehat{R}_N = 2r$, or $\widetilde{R}_N = r$. When plugging all the appropriate terms in (\ref{e32}) we get that the number of iterations until the evader region is  bounded by a circle with a radius that is less than or equal to $2r$ is,
\begin{equation}
N = \left\lceil {\frac{{\sqrt {{V_s}^2 - {V_T}^2} }}{{2\pi {V_T}}}\ln \left( {\frac{{r\left( {2 - {e^{\frac{{2\pi {V_T}}}{{\sqrt {{V_s}^2 - {V_T}^2} }}}}} \right)}}{{{R_0}\left( {1 - {e^{\frac{{2\pi {V_T}}}{{\sqrt {{V_s}^2 - {V_T}^2} }}}}} \right) + r{e^{\frac{{2\pi {V_T}}}{{\sqrt {{V_s}^2 - {V_T}^2} }}}}}}} \right)} \right\rceil 
\label{e33}
\end{equation} 
The difference equation for the cycle times are given by,
\begin{equation}
{T_{i + 1}} = {c_2}{T_i} + {c_3} 
\label{e34}
\end{equation}
Where in this context a cycle is defined as a traversal of an angle of $2\pi$ by the sweeper. The coefficient $c_3$ is given by,
\begin{equation}
{c_3} = \frac{{ - r\left( {{e^{\frac{{2\pi {V_T}}}{{\sqrt {{V_s}^2 - {V_T}^2} }}}} - 1} \right)}}{{{V_T}}}
\label{e35}
\end{equation}
The total sweep time until the evader region is bounded by a circle with a radius that is equal to or smaller than $2r$ is developed in Appendix $C$ of \cite{francos2019search} and is given by,
\begin{equation}
{T_{spiral}} = \frac{{{T_0} - {c_2}{T_{N - 1}} + \left( {N - 1} \right){c_3}}}{{1 - {c_2}}}
\label{e36}
\end{equation}
Where $T_0={t_{2\pi }}$ is the time of the first sweep given by,
\begin{equation}
{T_0} = \frac{{\left( {{R_0} - r} \right)\left( {{e^{\frac{{2\pi {V_T}}}{{\sqrt {{V_s}^2 - {V_T}^2} }}}} - 1} \right)}}{{{V_T}}}
\label{e37}
\end{equation}
The time it takes to complete the last cycle before the evader region is bounded by a circle with a radius that is smaller or equal to $2r$ is developed in Appendix $D$ of \cite{francos2019search} and is given by,
\begin{equation}
{T_{N - 1}} = \frac{{{c_3}}}{{1 - {c_2}}} + {c_2}^{N - 1}\left( {{T_0} - \frac{{{c_3}}}{{1 - {c_2}}}} \right)
\label{e38}
\end{equation}
Substitution of coefficients in (\ref{e38}) results in, 
\begin{equation}
{T_{N - 1}} = \frac{r}{{{V_T}}} + \frac{{{e^{\frac{{2\pi {V_T}\left( {N - 1} \right)}}{{\sqrt {{V_s}^2 - {V_T}^2} }}}}}}{{{V_T}}}\left( {{R_0}\left( {{e^{\frac{{2\pi {V_T}}}{{\sqrt {{V_s}^2 - {V_T}^2} }}}} - 1} \right) - r{e^{\frac{{2\pi {V_T}}}{{\sqrt {{V_s}^2 - {V_T}^2} }}}}} \right)
\label{e39}
\end{equation}
Thus (\ref{e36}) can be written as,
\begin{equation}
\begin{array}{l}
{T_{spiral}} =  - \frac{{{R_0}}}{{{V_T}}} - \frac{{r{e^{\frac{{2\pi {V_T}}}{{\sqrt {{V_s}^2 - {V_T}^2} }}}}}}{{{V_T}\left( {1 - {e^{\frac{{2\pi {V_T}}}{{\sqrt {{V_s}^2 - {V_T}^2} }}}}} \right)}} + \frac{{{e^{\frac{{2\pi {V_T}N}}{{\sqrt {{V_s}^2 - {V_T}^2} }}}}{R_0}}}{{{V_T}}} + \frac{{r{e^{\frac{{2\pi {V_T}\left( {N + 1} \right)}}{{\sqrt {{V_s}^2 - {V_T}^2} }}}}}}{{{V_T}\left( {1 - {e^{\frac{{2\pi {V_T}}}{{\sqrt {{V_s}^2 - {V_T}^2} }}}}} \right)}} \\ + \frac{{rN}}{{{V_T}}}
\end{array}
\label{e40}
\end{equation}

An interesting analysis is to determine the implications of having different ratios between sensor length and the initial evader region's radius on the number of required iterations and sweep times. We express $R_0$ as,
\begin{equation}
R_0 = \alpha r \hspace{1mm}, \hspace{1mm} \alpha  \geq 2
\label{e41}    
\end{equation}
In Fig. $5$ we observe the number of iterations and sweep times for $2\leq \alpha \leq 100$.
\begin{figure}[ht]
\noindent \centering{}\includegraphics[]{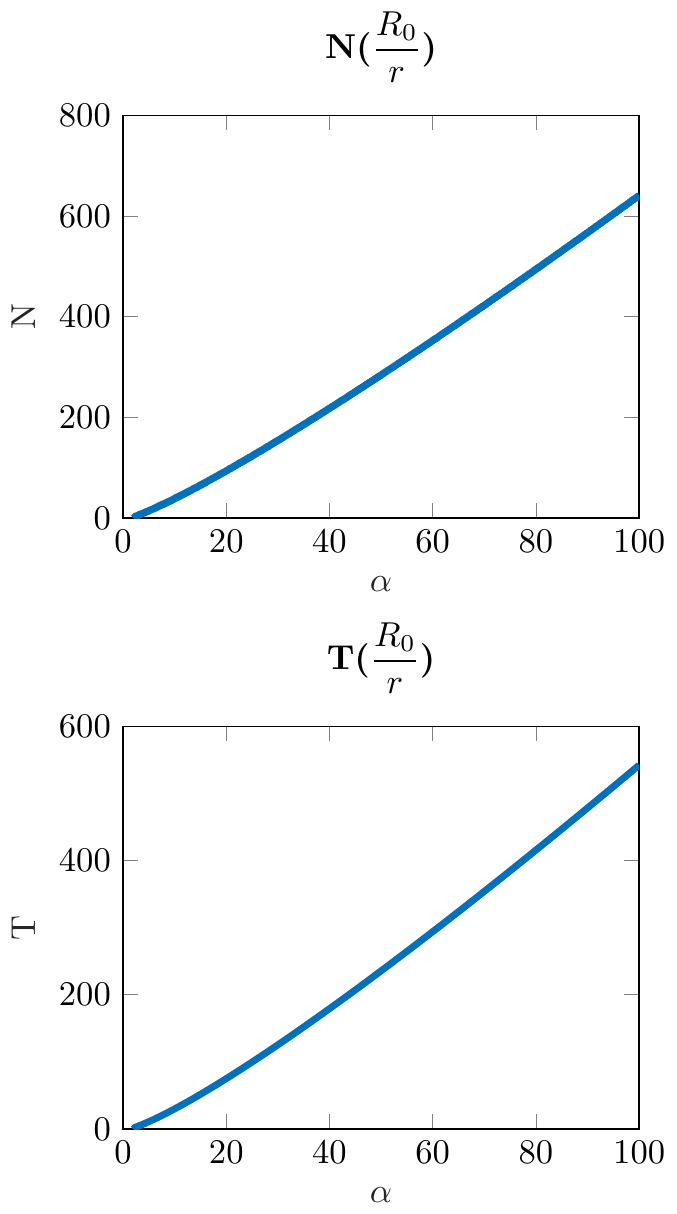} \caption{Number of iterations and sweep times for $2\leq \alpha \leq 100$ for a line formation of sweepers implementing the drifting spiral sweep protocol. The chosen values of the parameters are $V_T = 1$ and $\Delta V = V_T$.}
\label{Fig5Label}
\end{figure}

From Fig. $5$ we can view that as $\alpha$ increases, both $N$ and $T$ increase almost linearly. This is not intuitive when observing the equations derived for $N$ and $T$.
Another interesting analysis is to determine the implications of having different sweeper speeds on the number of sweeps and search times.
\begin{figure}[ht]
\noindent \centering{}\includegraphics[]{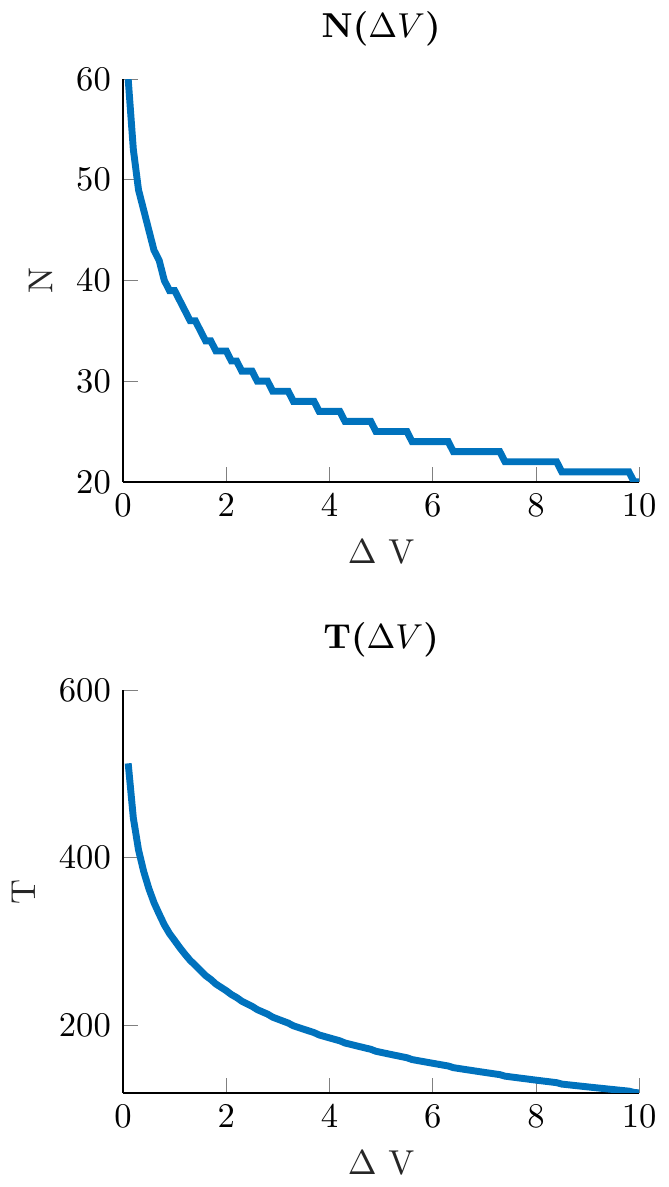} \caption{Number of iterations and sweep times for different choices of $\Delta V$ for a line formation of sweepers implementing the drifting spiral sweep protocol until the formation reduces the evader region to be bounded within a circle having a radius smaller than $2r$. In this figure $\Delta V$ varies between $0.1V_T\leq \Delta V \leq 10V_T$. The other chosen values of the parameters are $V_T = 1$, $R_0=100$ and $r=10$.}
\label{Fig6Label}
\end{figure}
From Fig. $6$ we observe that as $\Delta V$ increases, both $N$ and $T$ decrease in a piecewise exponential manner. This result is anticipated since the number of iterations must be an integer number. This implies that for nearly equal values of $\Delta V$, the number of iterations stays the same. Only when $\Delta V$ is sufficiently different, it will result in the decrease of the number of iterations by a number that is greater than $1$ iteration, and hence require a smaller number of sweeps to be performed in order to complete the mission.  Only when this situation occurs, it will be apparent in the plot.

\subsection{The End Game}
To facilitate detection of all evaders in the evader region, the sweeper formation must change its sweeping pattern once the evader region is reduced to within a circle of radius $2r$. The need to change the sweeping pattern arises since smart evaders that are located near the center of the evader region may move with very high angular speeds compared to the formation's angular speed. This constraint is given by the equations, ${\omega _s} = \frac{{{V_s}}}{2r}$ and ${\omega _T} = \frac{{{V_T}}}{\varepsilon }$. The first provides the line formation's angular speed and the second the evader's angular speed. Since $\varepsilon$ may be very small and approach zero, a smart evader may choose to trail behind the formation's sensor and thus never be detected. At the last sweep the evader region is bounded by a circle with a radius that is smaller than or equal to $2r$. 

Hence, to completely clean the evader region, the sweeper formation needs to change its sweeping protocol. The depiction of the scenario at the beginning of the end game is shown in Fig. $7(a)$. Following the last sweep, the sweeper formation moves upwards until the lower tip of its sensor is placed at the center of the evader region. During this upward movement the evader region continues to spread and the sweepers have to take this spread into account. The depiction after the sweeper formation completes the movement can be seen in Fig. $7(b)$. Therefore, the distance the formation advances is given by,
\begin{equation}
{r_{out}} = \frac{{\left( {2r - {R_N}} \right){V_s}}}{{{V_s} + {V_T}}}    
\label{e42}    
\end{equation}
The time it takes to perform this movement is given by, 
\begin{equation}
{T_{out}} = \frac{{2r - {R_N}}}{{{V_s} + {V_T}}}
\label{e43}    
\end{equation}
The expression for $R_N$ is derived in Appendix $B$ of \cite{francos2019search} and is given by,
\begin{equation}
\widetilde{R}_N = \frac{{{c_1}}}{{1 - {c_2}}} + {c_2}^N\left( {\widetilde{R}_0 - \frac{{{c_1}}}{{1 - {c_2}}}} \right)
\label{e44}
\end{equation}
Substitution of coefficients and writing the expression for $R_N$ instead of $\widetilde{R}_N$ yields,
\begin{equation}
\begin{array}{l}
{R_N} = \\ - \frac{{2r}}{{1 - {e^{\frac{{2\pi {V_T}}}{{\sqrt {{V_s}^2 - {V_T}^2} }}}}}} + {e^{\frac{{2\pi {V_T}N}}{{\sqrt {{V_s}^2 - {V_T}^2} }}}}\left( {\frac{{{R_0}\left( {1 - {e^{\frac{{2\pi {V_T}}}{{\sqrt {{V_s}^2 - {V_T}^2} }}}}} \right) + r\left( {1 + {e^{\frac{{2\pi {V_T}}}{{\sqrt {{V_s}^2 - {V_T}^2} }}}}} \right)}}{{1 - {e^{\frac{{2\pi {V_T}}}{{\sqrt {{V_s}^2 - {V_T}^2} }}}}}}} \right) + r
\end{array}
\label{e45}
\end{equation}
Following this outwards advancement, the sweeper formation performs an additional spiral sweep when the center of the formation is at a distance of $r$ from the center of the evader region. The time it takes to complete this sweep is denoted by $T_l$ and is given by,
\begin{equation}
{T_l} = \frac{{r\left( {{e^{\frac{{2\pi {V_T}}}{{\sqrt {{V_s}^2 - {V_T}^2} }}}} - 1} \right)}}{{{V_T}}}
\label{e46}
\end{equation}
During the last spiral sweep, the evader region spreads from its center point to a circle with a radius of,  
\begin{equation}
{R_{last}} = {T_l}{V_T} = r\left( {{e^{\frac{{2\pi {V_T}}}{{\sqrt {{V_s}^2 - {V_T}^2} }}}} - 1} \right)
\label{e47}
\end{equation}
The depiction of the scenario after the last spiral sweep is shown in Fig. $7(c)$. Following this last spiral sweep, the line formation advances downwards a distance of,
\begin{equation}
{R_{down}}=\frac{{\left( {r + \frac{{R_{last}}}{2}} \right){V_s}}}{{ {{V_s} + {V_T}} }}
\label{e48}
\end{equation}
So that there are equal lengths of the evader region around the two sensor tips. This time is given by, $T_{in\_last}=\frac{R_{down}}{V_s}$. After this downwards motion, the evader region is bounded by a circle with a radius of,
\begin{equation}
{R_f} = T_{in\_last}V_T + {R_{last}}
\label{e49}
\end{equation}
The depiction of the scenario at this time instance is shown in Fig. $7(d)$. Following this movement, the sweeper formation performs a linear motion that is similar to the linear motion that is performed in the end game of \cite{francos2019search} and completes the search protocol. An illustration of the $4$ stages of the end game are shown in Fig. $7$.
\begin{figure}
\noindent \centering{}\includegraphics[width=4.2in,height =4in]{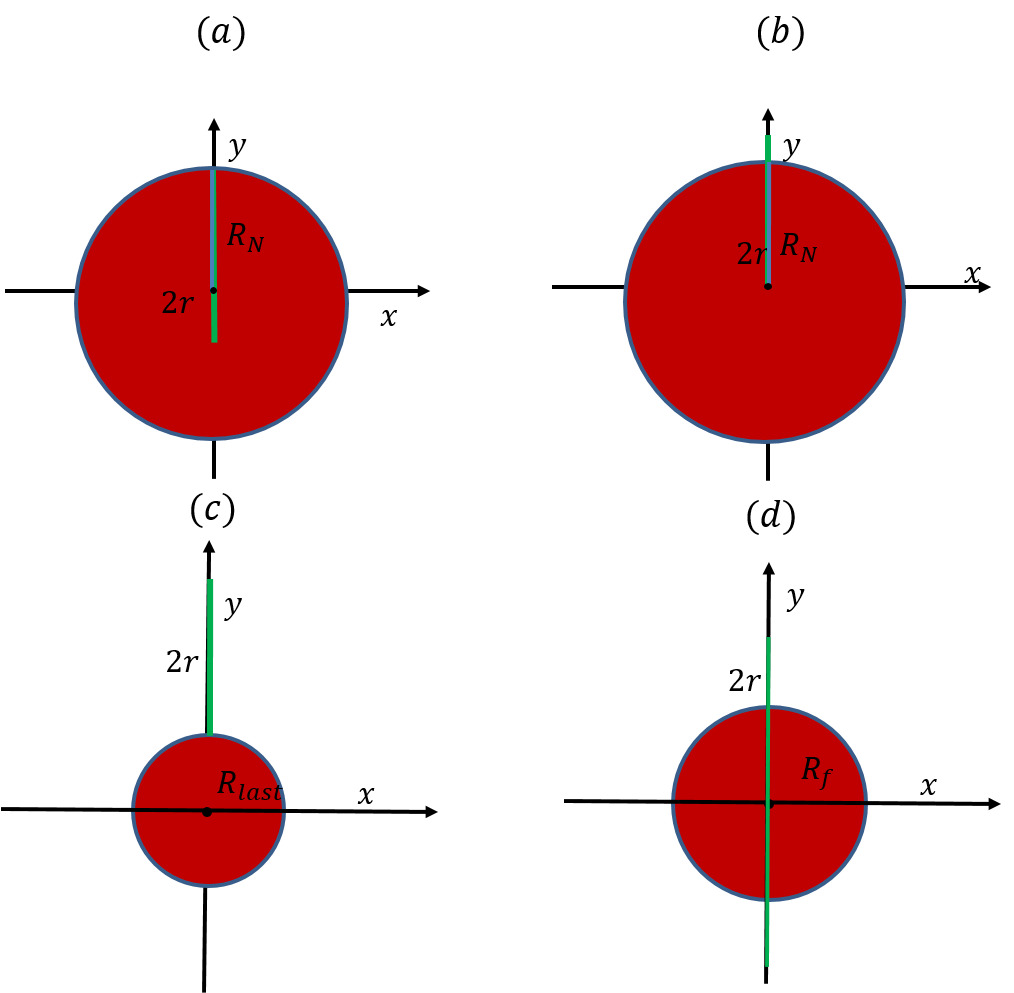} \caption{Different stages of the end game. The sweepers' sensors are shown in green. Red areas indicate additional locations where potential evaders may still be located. (a) - Depiction of the scenario at the beginning of the end game. (b) - Depiction of the scenario after the sweeper moves up and places the lower tip of its sensor at the center of the evader region. (c) - Depiction of the scenario after the last spiral sweep. (d) Depiction of the scenario prior to the linear sweep.}
\label{Fig7Label}
\end{figure}
In order for the described end-game maneuvers to be applicable and allow detection of all evaders, the margins between the tips of the formation's sensor in each direction and the evader region boundaries need to conform to following inequality,
\begin{equation}
\frac{{r - {R_{f}}}}{{{V_T}}} > {T_{linear}}
\label{e50}
\end{equation}
Let ${T_{linear}}$ denote the total time required for the sweeping formation to detect all evaders in the remaining right section of evader region along with the time required by the formation to sweep from the rightmost point it reaches up to the leftmost point the evader region expands to. These times are given by $t$ and $ \tilde t$. Hence, 
\begin{equation}
{T_{linear}} = t + \tilde t
\label{e54}
\end{equation}
For the described one dimensional right and left sweeps to be valid and guarantee detection of all evaders and completion of the mission, (\ref{e50}) has to hold, implying,
\begin{equation}
\frac{{r - {R_{f}}}}{{{V_T}}} > \frac{{{R_{f}}\left( {3{V_s} - {V_T}} \right)}}{{{{\left( {{V_s} - {V_T}} \right)}^2}}}
\label{e56}
\end{equation}
From substitution of the expressions for $V_s$ and ${R_{f}}$, (\ref{e56}) can be written as,
\begin{equation}
{V_s}^2\left( {r - {R_f}} \right) - {V_s}{V_T}\left( {2r + {R_f}} \right) + r{V_T}^2 > 0  
\label{e210}    
\end{equation}
The quadratic equation in (\ref{e210}) has $2$ positive roots. Therefore, in order for the one dimensional linear scan to be valid $V_s$ has to be greater than the largest positive root. Implying that,
\begin{equation}
{V_s} \geq \frac{{{V_T}\left( {2r + {R_f} + \sqrt {8r{R_f} + {R_f}^2} } \right)}}{{2\left( {r - {R_f}} \right)}}
\label{e211}    
\end{equation}
For a parameter choice of $R_0=100$, $r=10$, $V_T=1$, $\Delta V =1$ which leads to $R_f = 1.2629$, $V_s$ has to satisfy $V_s> 1.7966$ which clearly holds. 

The evader region's rightmost point of expansion spreads from $(R_{f},0)$ and spreads at a speed of $V_T$. Hence, if the constraint in (\ref{e50}) holds, then the rightwards and leftwards linear sweeps may be considered as a one dimensional sweep. Therefore, the time $t$ it takes the formation to clean the spread of potential evaders from the right section of the region can be calculated from, ${V_s}t = {R_{f}} + {V_T}t$. Therefore, $t$ is given by,
\begin{equation}
t = \frac{{{R_{f}}}}{{{V_s} - {V_T}}}.
\label{e51}
\end{equation}
$\tilde t$ is the time required for the linear formation when it is in $(tV_s,0)$ to change its sweeping direction and perform a leftwards sweep to a point that spread at a speed of $V_T$ from the leftmost point in the evader region at the origin of the search, the point $(-R_{f},0)$, for a time of ${\tilde t + t}$. Therefore, 
\begin{equation}
- {R_{f}} - {V_T}\left( {\tilde t + t} \right) = t{V_s} - {V_s}\tilde t
\label{e52}
\end{equation}
Substituting $t$ into the equation yields,
\begin{equation}
\tilde t = \frac{{2{V_s}{R_{f}}}}{{{{\left( {{V_s} - {V_T}} \right)}^2}}}
\label{e53}
\end{equation}

For $R_0 = 100$, $r=10$, $V_T = 1$, $\Delta V = 1$, $V_s=60.6435$, $t=0.0212, \tilde t=0.0431$ and ${T_{linear}} =0.0642$. Therefore, the total scan time until a complete cleaning of the evader region is given by,
\begin{equation}
T_{total} = T_{spiral} + {T_{out}} + T_l+ T_{in\_last} + T_{linear}
\label{e55}
\end{equation}
Which yields for the chosen values of the parameters that $T_{total} = 301.102 + 0.3066 + 1.0918 + 0.1445 + 0.0629= 302.7078$.  Fig. $8$ shows the total sweep time required to complete the detection of all evaders as a function of the line formation's speed above the drifting spiral critical speed.

\begin{figure}
\noindent \centering{}\includegraphics[]{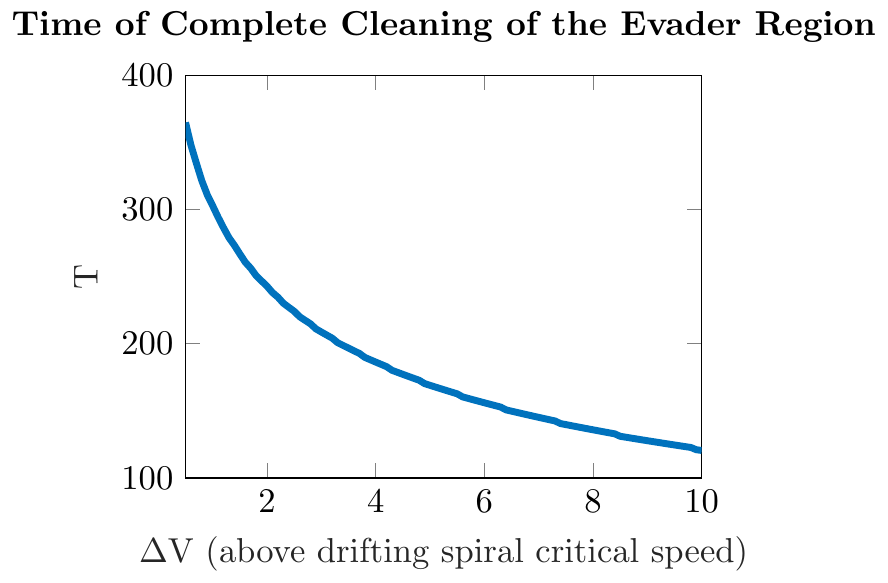} \caption{Sweep times until complete cleaning of the evader region. In this figure the line formation of sweepers that employ the drifting spiral sweep protocol moves with speeds that are $\Delta V$ above the drifting spiral protocol's critical speed. In this plot $\Delta V$ varies between $0.5V_T\leq \Delta V \leq 10V_T$. The other parameters values chosen for this plot are $V_T = 1$, $R_0=100$ and $r=10$.}
\label{Fig8Label}
\end{figure}

\section{The Improved Spiral Sweep Protocol}
\label{sec5}
Since our aim is to provide a sweep protocol that improves both the circular and drifting spiral sweep protocols we would like the sweepers to employ a more efficient motion throughout the sweep protocol. Similarly to the idea presented in the previous section, we wish that throughout the motion of the line formation of sweepers, its sensor's footprint will maximally overlap the evader region. An illustration of the initial placement of a line formation of agents that employ the proposed spiral sweep protocol is presented in Fig. $2$. It is the same as the initial placement of sweepers employing the drifting spiral sweep protocol. The line formation of sweepers has a combined sensor length of $2r$ and we desire the sweepers to have the lowest possible critical speed, hence we propose a different search protocol. The line formation of sweepers starts with a sensor length of $2r$ inside the evader region. If the formation's speed exceeds the required critical speed of the given scenario, the line formation reduces the evader region's area after completing a full sweep around it. The line formation of sweepers begins its spiral traversal with the upper tip of its sensor tangent to the boundary of the evader region at point $P =(0,R_0)$. In order to keep its sensor tangent to the evader region, the formation must travel at angle $\phi$ to the normal of the evader region. $\phi$ is calculated in equation (\ref{e3}) in the previous section. This method of traveling at angle $\phi$ preserves the evader region's circular shape. 

The equations of motion are calculated with respect to the center of the sweeper formation, denoted as in the previous section by,  $C\left( {{t_\theta }} \right)$. $C\left( {{t_\theta }} \right)$ describes the trajectory of the center of the line formation as a function of $\theta \left( {{t_\theta }} \right)$, the angle of the formation's sensor with respect to the center of the evader region. $\theta \left( {{t_\theta }} \right)$ describes the angle the formation travelled around the region and is calculated in the previous section in equation (\ref{e12}). ${t_\theta }$ denotes the sweep time after the formation completes a traversal of angle $\theta$ around the center of the evader region. The time it takes the line formation to travel an angle of $\theta$ around the center of the evader region, ${t_\theta }$ is calculated in the previous section in equation (\ref{e13}).

The first modification in the proposed spiral protocol with respect to the drifting spiral sweep protocol is that the sweeper formation travels more than a full sweep of $2\pi$ at each iteration around the evader region in order to detect all escaping smart evaders. The additional angle, denoted by $\beta$, needs to be traversed in order to detect all evaders that may have spread from the "most dangerous point" at the beginning of the sweep. Let us denote the most "most dangerous point" during the first sweep as point $P$. The angle $\beta$ depends on the radius of the circle that bounds the evader region. Hence the notation $\beta_0$ refers to the angle $\beta$ during the first sweep in which the evader region is bounded by a circle of radius $R_0$.

Point $P$ is adjacent to the upper tip of the sweeper formation's sensor. The proof that point $P$ is the "most dangerous point" follows similar steps as the proof of Theorem $2$ in \cite{francos2019search}, with the exception that since the sweeper formation travels in a spiral trajectory for an of $2\pi + \beta_0$ at each sweep, and only then advances inwards towards the center of the evader region, this point is shifted by $\beta_0$ after each sweep.

After the formation traverses the additional angle $\beta_0$, the evader region's boundary is due to spread of evaders that resided at the lower tip of the sensor during its sweep. The time it takes the sweeper formation to travel an angle of $2\pi  + {\beta _0}$, where ${\beta _0}$ is shown in Fig. $9$ is given by,   
\begin{equation}
{t_{2\pi  + {\beta _0}}} = \frac{{\left( {{R_0} - r} \right)\left( {{e^{\frac{{\left( {2\pi  + {\beta _0}} \right){V_T}}}{{\sqrt {{V_s}^2 - {V_T}^2} }}}} - 1} \right)}}{{{V_T}}}
\label{e309}
\end{equation}

\begin{figure}[ht]
\noindent \centering{}\includegraphics[width=2in,height =2in]{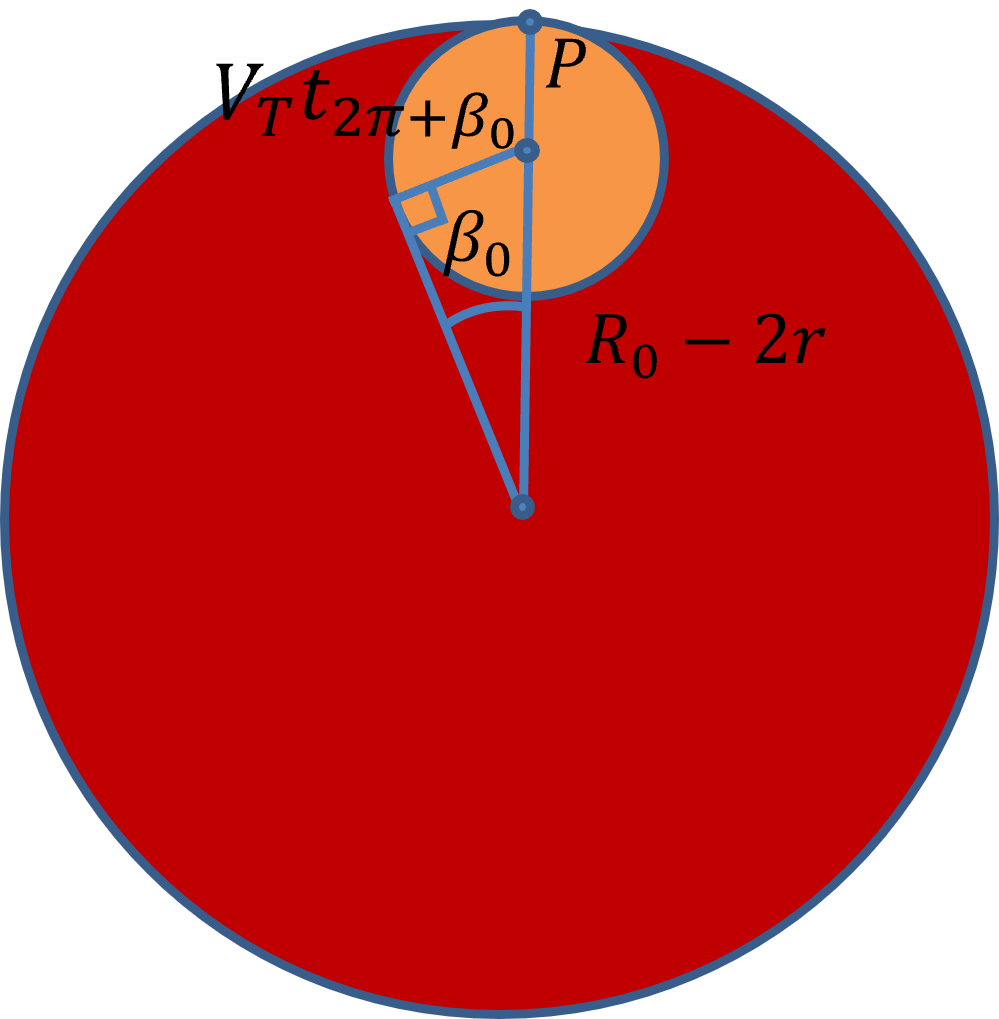} \caption{Geometric representation required for critical speed calculation. Red areas indicate locations where potential evaders may be located. The orange circle shows the spread of potential evaders to a circle of radius ${V_T}{t_{2\pi  + {\beta _0}}}$ from the lower tip of the formation's sensor.}
\label{Fig9Label}
\end{figure}

As mentioned earlier, the subscript $0$ in ${\beta _0}$ denotes the iteration or cycle number, indicating that the value of $\beta$ changes as the sweep protocol progresses. After the sweeper formation completes a traversal of $2\pi  + {\beta _0}$ around the evader region it moves towards the center of the evader region. During this motion its lower tip points to the center of the region. $\beta _0$ is given by,
\begin{equation}
\sin {\beta _0} = \frac{{{V_T}{t_{2\pi  + {\beta _0}}}}}{{{R_0} - 2r}}
\label{e310}
\end{equation}
The critical speed that satisfies the confinement task is computed numerically using the Newton method. When the sweepers travel towards the center of the evader region after completing a spiral sweep they have to meet the evader wavefront travelling outwards the region with a speed of $V_T$ at the previous radius $R_0$. Therefore, the expansion of the evader region after the first sweep at time ${t_{2\pi  + {\beta _0}}}$, has to satisfy that,
\begin{equation}
{V_T}{t_{2\pi  + {\beta _0}}} \leq \frac{{2r{V_s}}}{{{V_s} + {V_T}}}
\label{e330}
\end{equation}
The critical speed is obtained when we have equality in (\ref{e330}). In order to calculate $\beta_0$ that is obtained when the formation moves at the critical speed, the expression of ${V_T}{t_{2\pi  + {\beta _0}}}$ in (\ref{e330}) is substituted with its equivalent expression in (\ref{e310}). Hence $\beta_0$ is, 
\begin{equation}
{\beta _0} = \arcsin \left( {\frac{{2r{V_s}}}{{\left( {{V_s} + {V_T}} \right)\left( {{R_0} - 2r} \right)}}} \right)
\label{e311}
\end{equation}
Substituting the expression for ${t_{2\pi  + {\beta _0}}}$ in (\ref{e330}), yields
\begin{equation}
\left( {{R_0} - r} \right)\left( {{e^{\frac{{\left( {2\pi  + {\beta _0}} \right){V_T}}}{{\sqrt {{V_s}^2 - {V_T}^2} }}}} - 1} \right) = \frac{{2r{V_s}}}{{{V_s} + {V_T}}}
\label{e317}
\end{equation}
In order to solve for $V_s$ we write (\ref{e317}) as,
\begin{equation}
F\left( {{V_s}} \right) = \frac{{2r{V_s}}}{{{V_s} + {V_T}}} - \left( {{R_0} - r} \right)\left( {{e^{\frac{{\left( {2\pi  + \arcsin \left( {\frac{{2r{V_s}}}{{\left( {{V_s} + {V_T}} \right)\left( {{R_0} - 2r} \right)}}} \right)} \right){V_T}}}{{\sqrt {{V_s}^2 - {V_T}^2} }}}} - 1} \right)
\label{e318}
\end{equation}
From (\ref{e318}) we find $V_s$ using the Newton iterative root finding method whose equation is given by,
\begin{equation}
{V_{{s_{n + 1}}}} = {V_{{s_n}}} - \frac{{F\left( {{V_{{s_n}}}} \right)}}{{\frac{{\partial F\left( {{V_{{s_n}}}} \right)}}{{\partial {V_{{s_n}}}}}}}
\label{e321}
\end{equation}
We choose as our initial estimate the lower bound on the sweeper speed given by,
\begin{equation}
{V_{{s_0}}} = \frac{{\pi {R_0}{V_T}}}{r}= V_{LB}
\label{e319}
\end{equation}
Hence, through the described iterative convergence we obtain a solution for $V_s$, which is the improved spiral protocol's critical speed. We denote this speed as ${V_{c_{spiral2}}}$. After converging to a solution we obtain that for a parameter choice of $R_0=100$, $r=10$, $V_T =1$,  ${V_{c_{spiral2}}}=33.4294$. This result is only slightly larger than the lower bound on the sweeper speed, $V_{LB}$ which equals $31.4159$. An illustrative simulation that demonstrates the confinement task when the sweeper formation employs the improved spiral sweep protocol is presented in Fig. $10$.

\begin{figure}
\noindent \centering{}\includegraphics[width=1.5in,height =1.5in]{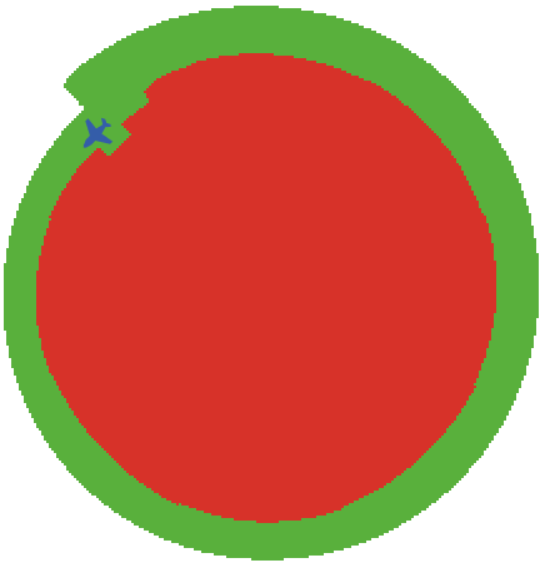} \caption{Swept areas and evader region status for a scenario where the line formation of agents successfully confines evaders to the initial evader region with the improved spiral sweep protocol. Green areas are locations that are free from evaders and red areas indicate locations where potential evaders may still be located.}
\label{Fig10Label}
\end{figure}

Denote by $\Delta V >0$ the addition to the formation's speed above the critical speed. The formation's speed is therefore given by, $V_s = {V_{c_{spiral2}}} + \Delta V$. If the sweeper formation moves with a speed greater than the critical speed, after each spiral sweep it can advance inwards towards the center of the evader region and sweep around an evader region that is bounded by a circle with a smaller radius. The total search time until the evader region is bounded by a circle with a radius that is less than or equal to $2r$ is given by the sum of the total spiral sweep times and the times of the inward advances. Namely,
\begin{equation}
T = {T_{in}} + {T_{spiral}}
\label{e328}
\end{equation}
After each iteration, the sweeper formation moves inwards towards the center of the evader region and the radius of the circle that bounds the region decreases. Consequently, the angle after which the formation moves inside after the next sweep changes as well. An illustrative simulation that demonstrates the cleaning progress of the evader region when the sweeper formation employs the improved spiral sweep protocol is presented in Fig. $11$. Green areas are locations that are free from evaders and red areas indicate locations where potential evaders may still be located.

\begin{figure}
\noindent \centering{}\includegraphics[width=2.4in,height =5.5in]{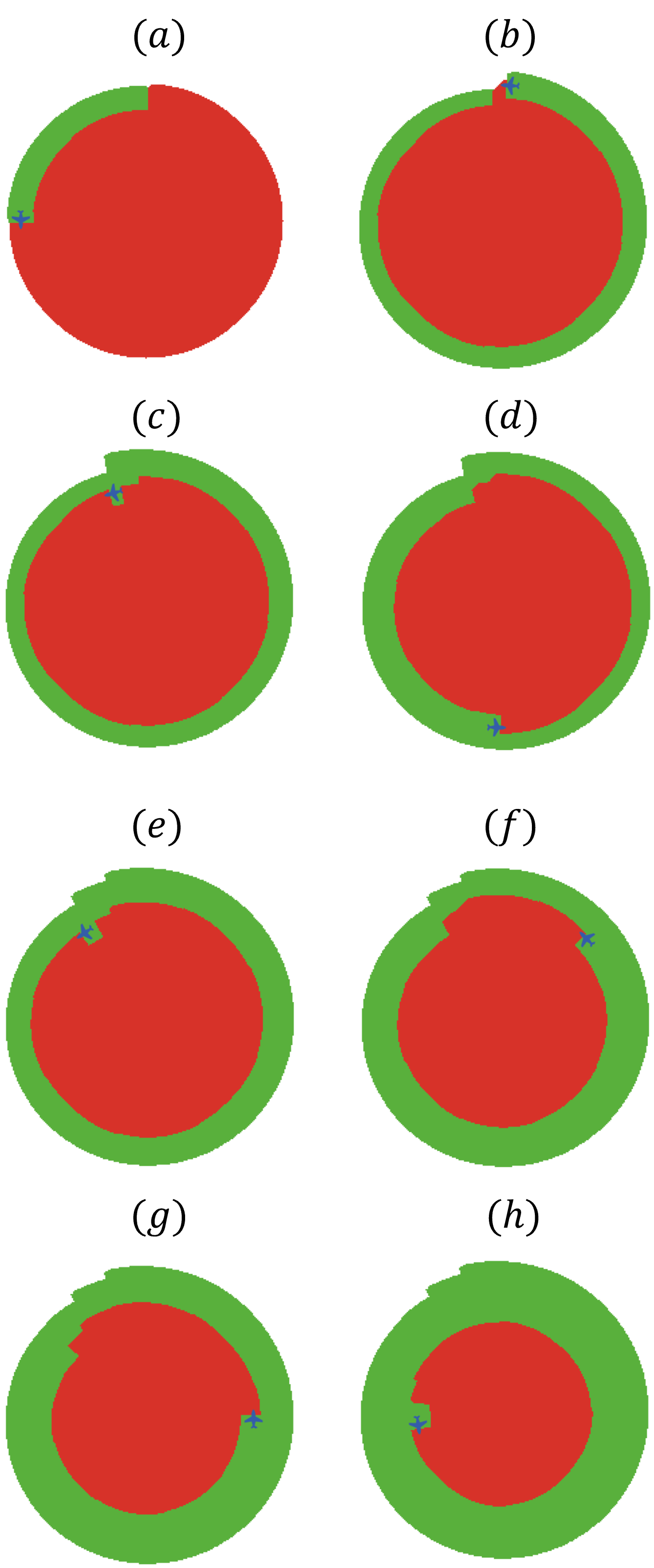} \caption{Swept areas and evader region status for different times in a scenario where the line formation of agents employs the improved spiral sweep protocol. The sweepers' sensors are shown in green. (a) - After a sweep by an angle of $\frac{\pi}{2}$. (b) - Towards the end of the first sweep. (c) - Beginning of the second sweep. (d) - Halfway through the second sweep. (e) - Beginning of the third sweep. (f) - Towards the end of the third sweep. (g) - After a sweep by an angle of $\frac{3\pi}{2}$ in the fourth sweep. (h) - Beginning of the sixth sweep. Green areas are locations that are free from evaders and red areas indicate locations where potential evaders may still be located.}
\label{Fig11Label}
\end{figure}

Therefore, after each sweep $\beta_i$ is calculated with respect to the new radius of the circle that bounds the evader region,
\begin{equation}
{\beta _i} = \arcsin \left( {\frac{{2r{V_s}}}{{\left( {{V_s} + {V_T}} \right)\left( {{R_i} - 2r} \right)}}} \right)
\label{e316}
\end{equation}
The time it takes to complete a spiral sweep of $2\pi +\beta_i$ around a region bounded by a circle of radius $R_i$ is given by, 
\begin{equation}
{T_{spira{l_i}}} = \frac{{\left( {{R_i} - r} \right)\left( {{e^{\frac{{\left( {2\pi  + {\beta _i}} \right){V_T}}}{{\sqrt {{V_s}^2 - {V_T}^2} }}}} - 1} \right)}}{{{V_T}}}
\label{e322}
\end{equation}
Given that the formation moves at a speed greater than the critical speed of the corresponding scenario we denote the distance the formation advances toward the center of the evader region by ${\delta _i}(\Delta V)$. In the term ${\delta _i}(\Delta V)$, $\Delta V$ denotes the increase in the formation's speed relative to the critical speed, and $i$ denotes the number of sweep iterations the sweeper formation performed around the evader region, where $i$ starts from sweep number $0$. This results in a new evader region bounded by a circle with a radius of ${R_{i + 1}} = {R_i} - {\delta _i}(\Delta V)$. We have that,
\begin{equation}
{\delta _i}(\Delta V) = 2r - {V_T}{T_{spiral{i}}}
\label{e324}
\end{equation}
As a function of the iteration number, we have that the distance the sweeper formation can advance inwards after completing an iteration is given by, 
\begin{equation}
{\delta _i}(\Delta V) = 2r - \left( {{R_i} - r} \right)\left( {{e^{\frac{{\left( {2\pi  + {\beta _i}} \right){V_T}}}{{\sqrt {{V_s}^2 - {V_T}^2} }}}} - 1} \right) \hspace{1mm}, \hspace{1mm} 0  \le {\delta _i}(\Delta V) \le 2r
\label{e325}
\end{equation}
After the sweeper formation finishes a full sweep, it moves inwards towards the center of the evader region while keeping the inner tip of its sensor directed to the center of the evader region. When the formation moves inwards, it does so with a speed of $V_s$, up to the position in which it starts its next sweep at the instance it meets the evader region's expanding wavefront. During the inwards advancements no detection of evaders occurs, while the evader region continues to expand.

The time required for the line formation to progress towards the center of the evader region until all of its sensor footprint overlaps the evader region depends on the relative speed between the formation's inwards entry speed and the evader region outwards expansion speed and is given in (\ref{e329}). As the formation moves in the direction of the evader region's center, the evader region continuous to expand. Hence, the formation must advance by a smaller distance than ${\delta _i}(\Delta V)$, denoted by ${\delta _{{i_{eff}}}}(\Delta V)$. This distance depends on the ratio between the sweeper formation inwards entry speed and the sum of the formation's entry and evader region's spread velocities (which are in the same direction). Therefore, ${\delta _{{i_{eff}}}}(\Delta V)$ describes the distance the sweeper formation travels after each sweep until it reaches the evader region's wavefront at the point where its entire sensor intersects the evader region. Hence, the distance the line formation can advance inwards after completing a sweep around the region is,
\begin{equation}
{\delta _{{i_{eff}}}}(\Delta V) = {\delta _i}(\Delta V)\left( {\frac{{{V_s}}}{{{V_s} + {V_T}}}} \right)
\label{e326}
\end{equation}
The evader region is therefore bounded by a circle with a radius given by,
\begin{equation}
{R_{i + 1}} = {R_i} - {\delta _i}(\Delta V)\left( {\frac{{{V_s}}}{{{V_s} + {V_T}}}} \right)
\label{e327}
\end{equation}
This protocol continues with a new calculation of $\beta_i$ for each iteration until the evader region is bounded by a circle with a radius that is smaller than $2r$. We denote this radius as $R_N$. Once the evader region is contained inside a circular domain with a radius of $2r < R_i < 4r$,  $\beta_i$ is,
\begin{equation}
{\beta _i} = \arcsin \left( {\frac{{\left( {{R_i} - 2r} \right){V_s}}}{{\left( {{V_s} + {V_T}} \right)\left( {{R_i} - 2r} \right)}}} \right)
\label{e312}
\end{equation}
The inwards advancement time depends on the iteration number. It is denoted by $T_{i{n_i}}$ and its expression is given by,
\begin{equation}
{T_{i{n_i}}} = \frac{{{\delta _{{i_{eff}}}}(\Delta V)}}{{{V_s}}} = \frac{{2r - \left( {{R_i} - r} \right)\left( {{e^{\frac{{\left( {2\pi  + {\beta _i}} \right){V_T}}}{{\sqrt {{V_s}^2 - {V_T}^2} }}}} - 1} \right)}}{{{V_s} + {V_T}}}
\label{e329}
\end{equation}
During the inward advancements only the tip of the sensor, that has zero width, is inserted into the evader region. Therefore, no evaders are detected until the sweeper formation completes its inward advance and starts sweeping again. After the sweeper line formation completes its advance into the evader region its sensor footprint over the evader region is equal to $2r$. For a parameter choice of $R_0=100$, $r=10$, $V_T=1$, $\Delta V =1$, $V_s= 34.4294$ the total time until reducing the evader region to be bounded by a circle of radius less than $2r$ is given by $T = T_{spiral} + T_{in}  = 222.0191 + 2.7655 = 224.7847$.

\subsection{The End Game}
In order to entirely clean the evader region, the sweeper formation needs to change the scanning method when the evader region is bounded by a circle of radius $2r$, due to the same consideration that are described in the end game of the drifting spiral sweep protocol. The depiction of the scenario at the beginning of the end game is shown in Fig. $12(a)$.
\begin{figure}
\noindent \centering{}\includegraphics[width=4in,height =3.6in]{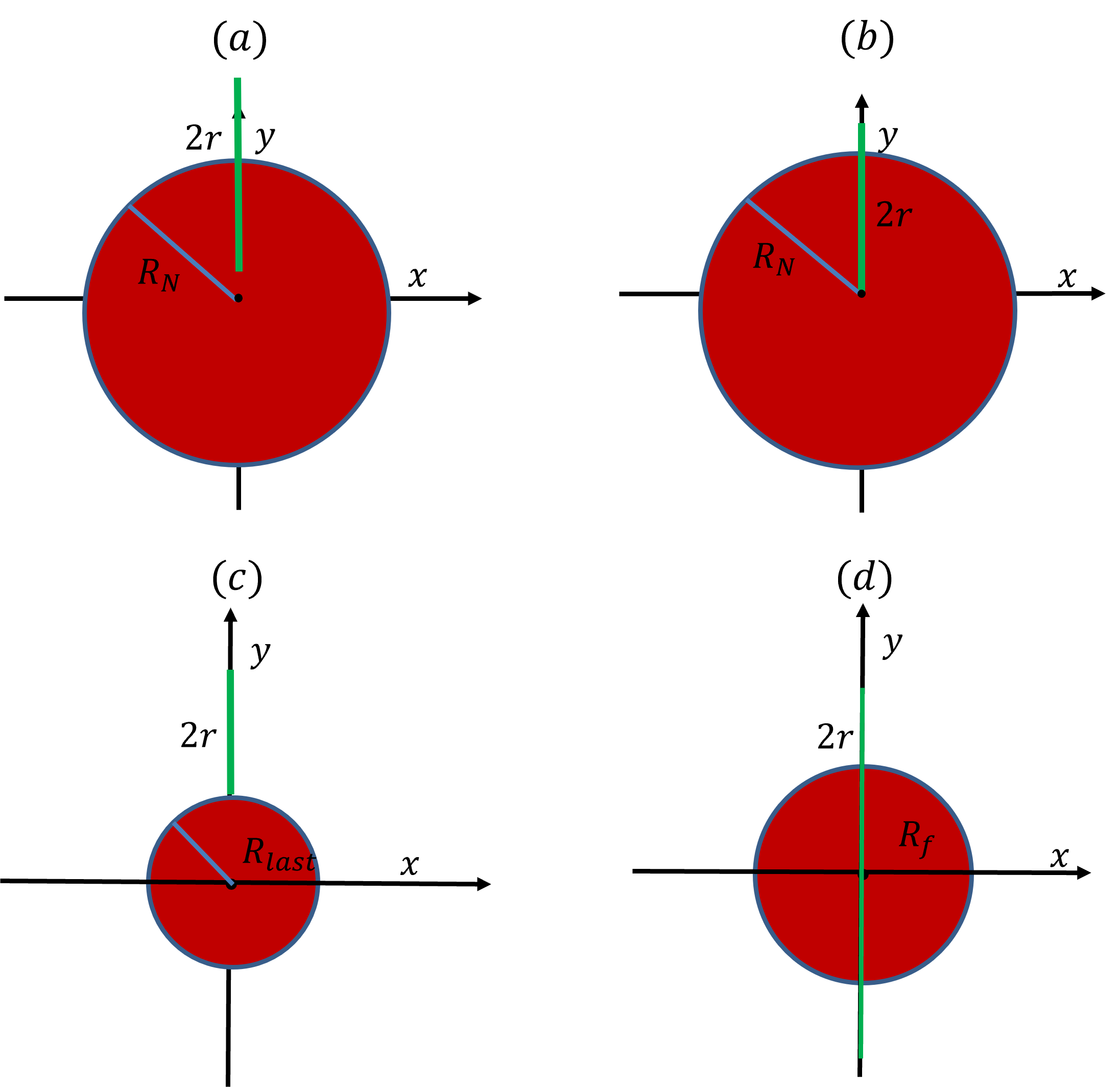} \caption{ Different stages of the end game. The sweepers' sensors are shown in green and the evader region is shown in red. (a) - Depiction of the state at the beginning of the end game. (b) - Depiction of the scenario after the sweeper moves towards the center of the evader region and places the lower tip of its sensor at the center of the region. (c) - Depiction of the scenario after the last spiral sweep. (d) - Depiction of the scenario prior to the linear sweep.}
\label{Fig12Label}
\end{figure}
In the last inward advancement, the sweeper formation places the tip of its sensor at the center of the evader region. The time it takes the formation to complete this movement is ${T_e} = \frac{{{R_N}}}{{{V_s}}}$.   

A parameter choice of $R_0=100$, $r=10$, $V_T=1$, $\Delta V = 1$, $V_s= 34.4294 $ yields that $T_e =0.139 $. Following the last inward advancement, the sweepers perform the last spiral sweep when the lower tip of the formation's sensor is placed at the center of the evader region as can be seen in Fig. $12(b)$. In order to apply a linear sweeping movement the last spiral sweep has to reduce the evader region to be bounded by a circle of radius less than $2r$. Following the last inwards advancement, the sweeper formation performs an additional spiral sweep when the center of the sweep formation is at a distance of $r$ from the center of the evader region. The time it takes to complete this sweep is denoted by $T_l$ and is given by,
\begin{equation}
{T_l} = \frac{{r\left( {{e^{\frac{{2\pi {V_T}}}{{\sqrt {{V_s}^2 - {V_T}^2} }}}} - 1} \right)}}{{{V_T}}}
\label{e78}
\end{equation}
The depiction after the sweeper formation completes this last spiral movement can be seen in Fig. $12(c)$. A parameter choice of $R_0=100$, $r=10$, $V_T=1$, $\Delta V =1$, $V_s=34.4294$ yields that $T_l =2.003 $. During the last spiral sweep, the evader region spreads from its center point to a circle with a radius of, 
\begin{equation}
{R_{last}} = {T_l}{V_T} = r\left( {{e^{\frac{{2\pi {V_T}}}{{\sqrt {{V_s}^2 - {V_T}^2} }}}} - 1} \right)
\label{e79}
\end{equation}
In order for a linear scan to be applicable ${R_{last}}$ has to be smaller than $2r$. This leads to a constraint,
\begin{equation}
r\left( {{e^{\frac{{2\pi {V_T}}}{{\sqrt {{V_s}^2 - {V_T}^2} }}}} - 1} \right) < 2r    
\label{e80}    
\end{equation}
yielding the requirement on the formation's speed,
\begin{equation}
{V_s} > {V_T}\sqrt {\frac{{4{\pi ^2}}}{{{{\left( {\ln 2} \right)}^2}}} + 1} 
\label{e82}    
\end{equation}
We previously observed that the spiral critical speed (for the considered spiral protocol) is close to the lower bound on the critical speed, $V_{LB}$, therefore we can check whether condition (\ref{e80}) is automatically satisfied. If the requirement on $V_s$ in (\ref{e82}) is less than $V_{LB}$, then, since the sweepers move with a speed above it, (\ref{e80}) will always be satisfied. Therefore if,
\begin{equation}
{V_T}\sqrt {\frac{{4{\pi ^2}}}{{{{\left( {\ln 2} \right)}^2}}} + 1}  < \frac{{\pi {R_0}{V_T}}}{r} = V_{LB}
\label{e83}    
\end{equation}
Or, if the ratio $\frac{{{R_0}}}{r}$ satisfies,
\begin{equation}
\sqrt {\frac{4}{{{{\left( {\ln 2} \right)}^2}}} + \frac{1}{{{\pi ^2}}}} \approx 2.9029  < \frac{{{R_0}}}{r}
\label{e84}    
\end{equation}
then ${R_{last}}$ is smaller than $2r$. 
Following this last spiral sweep, the line formation advances downwards in the direction of the center (and beyond it), a distance of,
\begin{equation}
{R_{down}} = \frac{{{V_s}\left( {r + \frac{{R_{last}}}{2}} \right)}}{{{V_s} + {V_T}}}
\label{e85}
\end{equation}
providing that there are equal lengths of the evader region around the two sensor tips. The time it takes the sweeper formation to complete this motion is given by,
\begin{equation}
{T_{down}} = \frac{{r + \frac{{R_{last}}}{2}}}{{{V_s} + {V_T}}}
\label{e86}    
\end{equation}
A parameter choice of $R_0=100, r=10, V_T=1, \Delta V =1, V_s= 34.4294 $ yields that $T_{down} = 0.3105$. During this time the evaders spread to a circle of radius $R_f$ around the center of the region given by,
\begin{equation}
{R_f} = {T_{down}}{V_T} + {R_{last}}
\label{e87}    
\end{equation}
The depiction of the scenario at this time instance is shown in Fig. $12(d)$. In order for the described end-game maneuvers to be applicable and allow detection of all evaders, the
margins between the tips of the formation’s sensor in each direction and the evader region
need to satisfy,
\begin{equation}
\frac{{r - {R_{f}}}}{{{V_T}}} > {T_{linear}}
\label{e88}
\end{equation}
This is the same constraint that appears in the end-game of the drifting spiral protocol as well. Following the end-game notations in the previous section of the drifting spiral protocol, ${T_{linear}}$ denotes the time it takes the linear formation to clean the right section of the remaining evader region in addition to the time it takes it to scan from the rightmost point it got to until the leftmost point of the expansion. These times are respectively denoted as $t$ and $ \tilde t$. Therefore, ${T_{linear}}$ is given by ${T_{linear}} = \tilde t + t$. In order to ensure the described one dimensional sweep is valid and consequently guarantees that no evader escapes during the completion of the detection mission, (\ref{e88}) must hold. Implying that,
\begin{equation}
{V_s} \geq \frac{{{V_T}\left( {2r + {R_f} + \sqrt {8r{R_f} + {R_f}^2} } \right)}}{{2\left( {r - {R_f}} \right)}}
\label{e97}    
\end{equation}
For a parameter choice of $R_0=100$, $r=10$, $V_T=1$, $\Delta V =1$ which leads to $R_f = 2.3135$, $V_s$ has to satisfy $V_s> 2.3491$ which clearly holds. 

Point $(R_{f},0)$ is the rightmost point from which the evader region spreads. As described in the end-game of the drifting spiral sweep protocol, in case the inequality in (\ref{e88}) holds, the rightwards and leftwards linear sweeps can be treated as a one dimensional sweep that enables the detection of all evaders. The time $t$ required for linear formation to detect all potential evaders spreading from the right section of the region can be calculated from, ${V_s}t = {R_{f}} + {V_T}t$. Hence, $t = \frac{{{R_{f}}}}{{{V_s} - {V_T}}}$.

$\tilde t$ is the time required for the line formation when it is in point $(tV_s,0)$ to change its sweeping direction and perform a leftwards sweep to a point that spread at a speed of $V_T$ from the leftmost point in the evader region at the origin of the search, the point $(-R_{f},0)$, for a time given by ${\tilde t + t}$. These results follow similar calculation as the ones in the end-game of the drifting spiral protocol and yields that
$\tilde t = \frac{{2{V_s}{R_{f}}}}{{{{\left( {{V_s} - {V_T}} \right)}^2}}}$. Recall that by definition ${T_{linear}} = t + \tilde t$. 
For $R_0 = 100$, $r=10$, $V_T = 1$, $\Delta V = 1$, $V_s= 34.4294 $, $t=0.0692$ , $\tilde t=0.1426$ and ${T_{linear}} =0.2118 $. Therefore, the total sweep time until complete detection of all evaders is,
\begin{equation}
T_{total} = T_{spiral} + T_{in} + T_e + {T_{down}} + T_l + T_{linear}
\label{e93}
\end{equation}
Which yields for the chosen values of the parameters that $T_{total} = 227.4489 $. Fig. $13$ shows the obtained sweep times until complete cleaning of the evader region when the line formation employs the improved spiral sweep protocol.

\begin{figure}[ht]
\noindent \centering{}\includegraphics[]{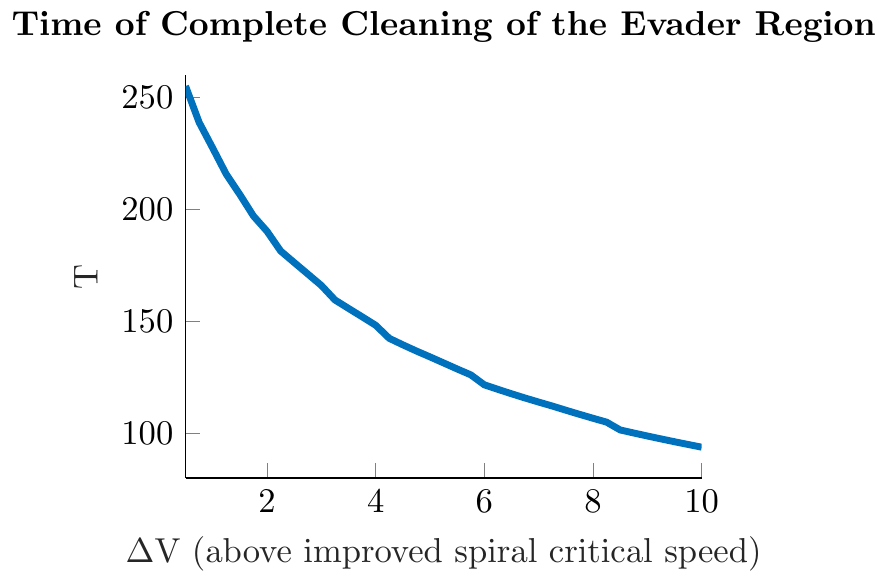} \caption{Sweep times until complete cleaning of the evader region. In this figure the line formation of sweepers that employ the improved spiral sweep protocol move with speeds that are $\Delta V$ above the improved spiral sweep protocol's critical speed. In this plot $\Delta V$ varies between $0.5V_T\leq \Delta V \leq 10V_T$. The other parameters values chosen for this plot are $V_T = 1$, $R_0=100$ and $r=10$.}
\label{Fig13Label}
\end{figure}

\section{Comparison of Line Formation Sweep Protocols}
\label{sec6}

In this section we compare the time it takes line formations of sweepers that employ spiral sweeps versus circular sweeps to clean the evader region completely. Since the critical speeds of the drifting and improved spiral sweep protocols are lower than the circular critical speed, a comparison of sweep times is performed when line formations that employ the different sweep protocols are performed with equivalent sensing capabilities and move at equal speeds above the circular critical speed. In Fig. $14$ we plot the times it takes a line formation of sweepers that employ the drifting spiral, improved spiral and the circular sweep protocols to clean the evader region when the sweeper formation moves with a speed of $\Delta V$ above the critical circular sweeper speed. The circular critical speed for the line formation scenario is derived in \cite{francos2019search} and is given by 
\begin{equation}
{V_c} = \frac{{2\pi {R_0}{V_T}}}{r} + V_T 
\label{e98}
\end{equation}
Fig. $14$ shows the reduction in sweep times that are obtained when using the spiral sweep protocols, and in particular the improved spiral sweep protocol favourable performance. If the user can choose for a given sweeper speed that is above the circular critical speed which sweep method to employ, the obvious choice is to sweep the area with a spiral protocol. It is to be noted that throughout the circular sweep protocol that is employed by the line formation of agents, as well as in the improved spiral protocol the center of the evader region does not change while in the drifting spiral protocol's setting the center of the region moves upwards as the search progresses. 
\begin{figure}[ht]
\noindent \centering{}\includegraphics[]{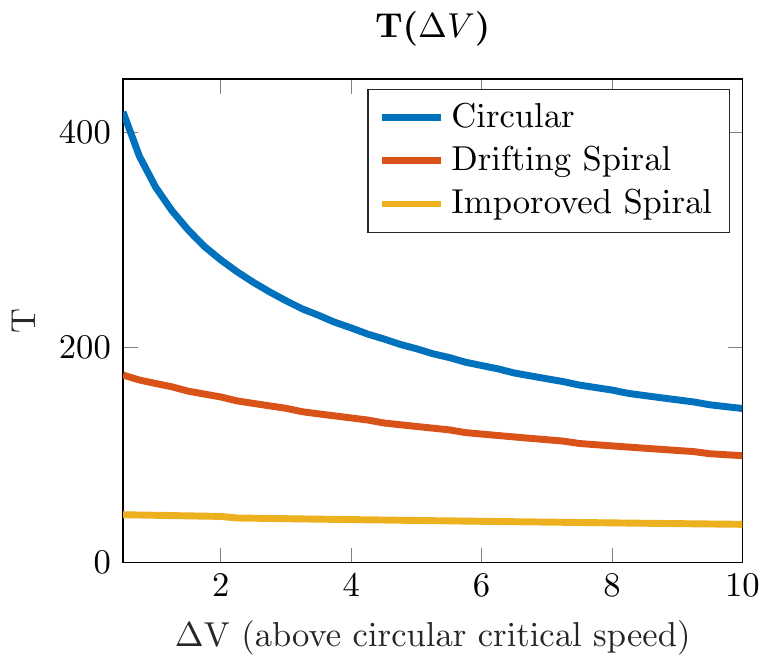} \caption{Sweep times until complete cleaning of the evader region. In this figure the line formation of sweepers that employ the spiral and circular sweep protocols move with speeds that are $\Delta V$ above the circular critical speed. In this plot $\Delta V$ varies between $0.5V_T\leq \Delta V \leq 10V_T$. The other parameters values chosen for this plot are $V_T = 1$, $R_0=100$ and $r=10$.}
\label{Fig14Label}
\end{figure}

From the results displayed in Fig. $14$ it is clear that performing the improved spiral protocol is clearly best since it provides considerably shorter sweep times. From the discussion in the previous sections of the paper and the resulting critical speeds for each of the protocols, we conclude that the improved spiral protocol requires a smaller critical speed compared to the circular and drifting spiral search protocols. Hence, with respect to both performance metrics performing the improved spiral protocol is best.  

\section{Conclusions}
\label{sec7}
We investigate a search problem where a line formation of sweeping agents (or a single agent with equivalent sensing capabilities), must guarantee detection of all smart evaders that aim to escape from a given circular domain while evading the sweeping agents. We propose $2$ types of spiral search protocols. The first protocol, the drifting spiral protocol, consists of only spiral sweeps. We provide an analysis on how to base the minimal speed a sweeper should have, in order to reduce the evader region to be bounded by a circle with a smaller radius than the sweeper formation's sensor length. We compare the developed critical speed to a lower bound on the critical speed that is independent of the search protocol the sweeper formation employs. In the drifting spiral protocol the center of the region moves upwards while the region shrinks until all evaders in the region are located. The second spiral protocol that is presented is an improved spiral sweep protocol around a fixed center that consists of spiral and inward linear motions. Unlike the drifting spiral protocol, it does not allow for a development of explicit formulas for the sweep time. For the improved spiral protocol we provide an analysis on how to base the minimal speed a sweeper should have and compare it to the lower bound as well. Both of the spiral sweep protocols yield a lower critical speed than the circular line formation critical speed developed in \cite{francos2019search}, and thus the spiral sweep protocols enable the sweeper line formation to scan faster and more efficiently a given region. Furthermore, sweeping with the spiral protocols also enables, given a limit on the sweepers speed, to sweep successfully larger regions compared to a line formation of agents that employs the circular sweep method.

We prove that smart evaders located near the center of the evader region can prevent the line formation of sweepers from detecting them if the formation moves only with a spiral sweep, and therefore after the evader region is reduced to be within a circle of radius $2r$ or less, a final end-game maneuver is performed to detect all evaders and succeed in the complete detection task. For the drifting spiral protocol the modification consists of a combination of an outwards advancement, an additional spiral sweep, a downwards motion and a final linear motion. For the improved spiral protocol the modification consists of placing the lower tip of the sensor at the center of the region, employing a final spiral sweep, moving downwards and then performing a final linear motion.

We show that with respect to both considered performance metrics, critical speed and search time, performing the improved spiral protocol is clearly best. We show that the improved spiral protocol's critical speed is only slightly larger than the theoretical lower bound on a sweeper's speed. We show that compared to the circular sweep protocol, the improved spiral protocol's search time requires a fraction of the search time and that is also considerably better than the drifting spiral protocol.

\section*{Declarations}
\begin{itemize}
\item Funding: This research was partially supported by the Technion Autonomous Systems Program (TASP).
\item Conflicts of interest/Competing interests: not applicable.
\item Availability of data and material: A video attachment is submitted with the manuscript as supplementary material.
\item Code availability: not applicable.
\item Authors' contributions: All authors contributed to the study conception and design. Analysis and development of methodology were performed by Roee M. Francos and Alfred M. Bruckstein. The manuscript was written by Roee M. Francos. Visualization of the results and the creation of software simulations were performed by Roee M. Francos. All authors read and approved the final manuscript.
\item Ethics approval: not applicable.
\item Consent to participate: not applicable.
\item Consent for publication: not applicable.
\end{itemize}


\begin{thebibliography}{00}

\bibitem{francos2019search}
R. M. Francos and A.M. Bruckstein, "Search for Smart Evaders with Sweeping Agents," {\it Robotica }{\bf 39}(12), 2210-2245  (2021).

\bibitem{mcgee2006guaranteed}
T. G. McGee and J. K. Hedrick, "Guaranteed Strategies to Search for Mobile Evaders in the Plane," {\bf In}: {\it Proceedings of the IEEE  American Control Conference}, (2006).
  
\bibitem{hew2015linear}
P. C. Hew, "Linear and Concentric Arc Patrols Against Smart Evaders," {\it Military Operations Research} {\bf20}(3), 39-48 (2015).


\bibitem{tang2006non}
T. Zhijun and U. Ozguner, "On Non-escape Search for a Moving Target by Multiple Mobile Sensor Agents," {\bf In}: {\it Proceedings of the IEEE  American Control Conference}, (2006).


\bibitem{tisue2004netlogo}
S. Tisue and U. Wilensky, "Netlogo: A simple environment for modeling complexity," {\bf In}: {\it Proceedings of the International conference on complex systems}, \textbf{21}, 16-21 (2004).

\bibitem{stone2}
L. D. Stone, J. O. Royset and A. R. Washburn, "Optimal Search for Moving Targets (International Series in Operations Research \& Management Science 237)" (Cham, Switzerland, Springer, 2016).

\bibitem{rekleitis}
I. Rekleitis, V. Lee-Shue, A.P. New and H. Choset, "Limited Communication, Multi-Robot Team Based Coverage," {\bf In}: {\it Proceedings of the IEEE International Conference on Robotics and Automation}, \textbf{4}, 3462-3468 (2004).

\bibitem{passino}
K. Passino, M. Polycarpou, D. Jacques, M. Pachter, Y. Liu, Y. Yang, M. Flint and M. Baum, "Cooperative Control for Autonomous Air vehicles," {\bf In}: {\it Cooperative Control and Optimization}, (Boston, MA, Springer, 2002) pp. 233-271. 


\bibitem{alpern2006theory}
S. Alpern and G. Shmuel, "The theory of search games and rendezvous,"  {\it Springer Science \& Business Media, 2006} {\bf55}, (2006). 



\bibitem{bertuccelli}
L. F. Bertuccelli and J. P. How, "Search for Dynamic Targets with Uncertain Probability Maps," {\bf In}: {\it Proceedings of the IEEE  American Control Conference}, (2006).

\bibitem{chung}
H. Chung, E. Polak, J.O. Royset and S. Sastry, "On the optimal detection of an underwater intruder in a channel using unmanned underwater vehicles,"{\it  Naval Research Logistics}  {\bf58}(8), 804-820 (2011).

\bibitem{koopman1980search}
  B. O. Koopman,  "Search and Screening: General Principles with historical applications," (New York, Pergamon Press, 1980).
 

\bibitem{vincent2004framework}
V. Patrick and I. Rubin,  "A framework and Analysis for Cooperative Search Using UAV Swarms," {\bf In}: {\it Proceedings of the ACM symposium on Applied computing}, (2004).  


\bibitem{altshuler2008efficient}
Y. Altshuler, V. Yanovsky, I. A. Wagner and A. M. Bruckstein, "Efficient Cooperative Search of Smart Targets Using UAV Swarms," {\it Robotica }{\bf 26}(4), 551-557 (2008).

\bibitem{wagner1997cooperative}
I. A. Wagner and A. M. Bruckstein, "Cooperative cleaners: A case of distributed ant-robotics,"
{\bf In}: {\it Communications, Computation, Control, and Signal protocoling: A Tribute to Thomas Kailath} (Kluwer Academic Publishers, The
Netherlands, 1997) pp. 289–308.

\bibitem{altshuler2005swarm}
Y. Altshuler, A. M. Bruckstein and I. A. Wagner, "Swarm robotics for a dynamic cleaning problem," {\bf In}: {\it Proceedings 2005 IEEE Swarm Intelligence Symposium}, (2005).


\bibitem{altshuler2011multi}
Y. Altshuler, V. Yanovsky, I. A. Wagner and A. M. Bruckstein, "Multi-agent Cooperative Cleaning of Expanding Domains," {\it The International Journal of Robotics Research} {\bf 30}(8), 1037-1071 (2011).


\bibitem{bressan1}
A. Bressan, "Differential inclusions and the control of forest fires," {\it Journal of Differential Equations} {\bf 243}(2), 179-207 (2007). 

\bibitem{bressan2}
A. Bressan, M. Burago, A. Friend and J. Jou, "Blocking Strategies for a Fire Control Problem," {\it Analysis and Applications}, {\bf 6}(03), 229-246 (2008).

\bibitem{bressan3}
A. Bressan, and C. De Lellis, "Existence of Optimal Strategies for a Fire Confinement Problem," {\it Communications on Pure and Applied Mathematics: A Journal Issued by the Courant Institute of Mathematical Sciences}, {\bf 62}(6), 789-830 (2009). 

\bibitem{bressan4}
A. Bressan and T. Wang, "On the Optimal Strategy for an Isotropic Blocking Problem," {\it Calculus of Variations and partial differential equations}, {\bf 45}(1-2), 125-145 (2012).

\bibitem{bressan5}
A. Bressan, "Dynamic Blocking Problems for a Model of Fire Propagation" {\bf In}: {\it Advances in Applied Mathematics, Modeling, and Computational Science}, (Boston, MA, Springer, 2013)
pp. 11-40.


\bibitem{Klein}
R. Klein, E. Langetepe, B. Schwarzwald, C. Levcopoulos and A. Lingas, "On a fire fighter's problem," {\it International Journal of Foundations of Computer Science} {\bf 30}(2), 231-246 (2019).

\bibitem{brown}
D. Brown and L. Sun, "Dynamic Exhaustive Mobile Target Search Using Unmanned Aerial Vehicles,"
{\it IEEE Transactions on Aerospace and Electronic Systems} {\bf 55}(6), 3413-3423 (2019).

\end{thebibliography}


\end{document}